%% file: main.tex
\documentclass[11pt]{article}
\usepackage[margin=1in]{geometry}
\usepackage[T1]{fontenc}
\usepackage{lmodern}
\usepackage{amsmath,amssymb,amsthm}
\usepackage{booktabs}
\usepackage{seqsplit}
\usepackage{enumitem}
\usepackage[numbers,sort&compress]{natbib}
\usepackage[hidelinks]{hyperref}
\usepackage{xcolor}
\usepackage{graphicx}
\usepackage{longtable}
\usepackage{placeins}
\usepackage{array}

\input{tables/meta.tex}
\input{tables/gear_macros.tex}

\newcommand{\inv}{\operatorname{inv}}
\newcommand{\RN}{\operatorname{RN}}
\newcommand{\ulp}{\operatorname{ulp}}
\newcommand{\code}[1]{\texttt{#1}}
\newtheorem{claim}{Claim}

\title{The involute function and its inverse in IEEE~754 double precision:\\
measured and proved accuracy, and its effect on gear inspection}

\author{Andrew Prestridge, PhD\thanks{B.S. 1985, M.S. 1991, Ph.D. 1996, Petroleum Engineering, Colorado School of Mines.
Member of MPMA (Motion + Power Manufacturers Alliance, formed by the merger of AGMA and ABMA); formerly an AGMA member.}\\
\small Evolvent Design, LLC (GearCalculators.com)\\
\small \href{mailto:andrew@evolventdesign.com}{\texttt{andrew@evolventdesign.com}}}

\date{November 2026}

\begin{document}
\maketitle

\begin{abstract}
Almost every spur and helical gear calculation uses the involute function $\inv(\alpha)=\tan\alpha-\alpha$ or its
inverse. Near $\alpha=0$ the two terms almost cancel: at $\alpha=10^{-3}$ the textbook formula in binary64 keeps about 9
correct digits, and far fewer in binary32, binary16 or fixed point, and the result can depend on the platform's
\code{tan}. We report the accuracy of a proprietary implementation that avoids these problems. On 50{,}000 committed
test angles the forward function's largest error against an 80-digit reference is $0.4999681$ ulp, so every tested
result is correctly rounded (measured, not proved). The inverse is proved faithfully rounded, with error at most
$0.6$ ulp, for $10^{-200}\le y<2$, using interval arithmetic and the SMT solvers z3 and cvc5 with a short, named list of
trusted inputs; its largest measured error on 50{,}000 points is $0.5555115$ ulp. A separate bracket around the root is proved to contain
it for $0<y\le10^6$, and did so at all 47{,}083 test points. Both functions return identical bits
on x86-64 and aarch64 Linux, Apple-silicon macOS, AWS Lambda and an NVIDIA DGX Spark, in Python and Node.js, on every
finite test input. Against public inverse methods on nine runtimes, our inverse returned the same bits on all of them
and stayed within $0.54$ ulp on 12{,}017 targets from $0.5^\circ$ to $60^\circ$, while textbook Newton iterations differed
between runtimes on up to about 10\% of inputs and kept only 12 correct digits. For external and internal spur gears
with 5 to 500 teeth and pressure angles from $14.5^\circ$ to $30^\circ$, all inputs fall inside the tested and proved
ranges. Against ISO~1328-1 tolerance classes, we give in micrometres the errors that shortcut inverses cause in pin
measurement and centre distance, and how far circular-arc CAD teeth (SolidWorks Toolbox) depart from the involute.
Drawing an involute is easy; accuracy matters in inspection.
\end{abstract}

\paragraph{In plain terms.} Every gear tooth is built on one small formula, $\inv(\alpha)=\tan\alpha-\alpha$. Gear
inspection also needs it run backwards. 

\section{Introduction}\label{sec:intro}

\subsection{Why the involute function matters}
Gears were made for centuries before anyone had a theory of their teeth. That theory began in the
eighteenth century. Camus's 1733 memoir to the Paris Academy~\cite{camus1733}, and the
gear books of his \emph{Cours de math\'ematique}~\cite{camus1806}, treated the condition for smooth tooth action; the
teeth he worked with were cycloidal~\cite{litvin1998}. Euler's first paper on tooth shape (1760) considered the involute of a circle and set it aside~\cite[\S\S19--22]{euler1760}. His second (1767) proposed it: if the angle between the
line of action and the line of centres is held constant, both tooth curves are involutes of circles about the wheel
centres, and these, being very easily drawn, ``seem to give the most convenient solution''~\cite[\S29]{euler1767}.
Savary later gave the curvature relation now called the Euler--Savary equation~\cite{volkov2014}.
Radzevich proposes calling the fundamental law of gearing the Camus--Euler--Savary (``CES'')
theorem~\cite{radzevich2019}. Practice followed. For circular-arc teeth, Willis's \emph{Principles of Mechanism} (1841) used a constant obliquity of $75^\circ30'$ to the line of centres, that is, a $14.5^\circ$ pressure angle~\cite{willis1841}, and Brown \& Sharpe's 1886 treatise drew lines of pressure at $75\tfrac12^\circ$ to the
radius~\cite{brownsharpe1886}. Today $20^\circ$ is the standard basic-rack angle~\cite{iso53,din867later}. The author's video
on the pressure angle tells this history~\cite{prestridge2023}; Appendix~\ref{app:history} gives a fuller, sourced
account. Nearly all cylindrical gears made today have involute teeth~\cite{litvin2004,iso21771}. If a
point on an involute profile has pressure angle $\alpha$ at radius $r$, its polar angle measured from the start of the
involute is $\inv(\alpha)=\tan\alpha-\alpha$, in radians, as ISO~21771-1:2024 defines it (its 4.4.5,
Formula~(11)). Gear calculations use the function both ways. The forward direction gives
tooth thickness at any diameter. The inverse, which finds $\alpha$ from $y=\inv(\alpha)$, gives the operating pressure angle of shifted gears or gears at a non-standard centre distance, the pressure angle at a measuring pin, and the centre distance for a given backlash. The inverse has no closed form. Gear programs usually solve it by iteration, and Cheng~\cite{cheng1992} gives a
widely used explicit approximation.

\subsection{Why it is hard to compute reliably}\label{sec:hard}
Three effects spoil a naive implementation.

\paragraph{Cancellation.} For small $\alpha$, $\tan\alpha=\alpha+\alpha^3/3+\cdots$. So $\tan\alpha$ and $\alpha$ share
their leading digits, and the subtraction removes them. If $\tan\alpha$ is correctly rounded, its error is up to
half an ulp of $\tan\alpha\approx\alpha$, while the result is about $\alpha^3/3$. So the relative error of the
difference grows by about $3/\alpha^2$, which is $3\times10^6$ at $\alpha=10^{-3}$. The loss comes from the subtraction, not from the math library: at the tested angles, a correctly rounded $\tan$ followed by one binary64 subtraction loses as many digits as the platform's $\tan$ (Table~\ref{tab:precision}).

\paragraph{Short formats.} Code written for embedded controllers, GPUs or machine-learning accelerators
often uses binary32, binary16 or fixed point~\cite{ieee754}. Each has fewer digits to start with, and the cancellation removes as many leading digits as it does in binary64. Fixed point is worse, because its resolution is absolute.
Once $\inv(\alpha)$ is smaller than the format's least significant bit, the result has no correct digit.

\paragraph{Platform dependence.} IEEE~754 requires correct rounding for $+,-,\times,/$ and $\sqrt{\ }$, but
not for $\tan$. System math libraries differ in the last bits of $\tan$, and measured worst-case errors vary
between libraries and versions~\cite{gladman2024}. Compilers may also contract $a\times b+c$ into a fused
multiply-add, and older x87 code evaluates in extended precision and rounds twice. A naive involute can
therefore return different bits on different machines, and the cancellation magnifies those differences. A
correctly rounded library such as CORE-MATH~\cite{coremath2022} removes the variation in $\tan$ itself but not
the cancellation.

A gear designer needs the same accurate answer from the same inputs on every machine. We report how well our $\inv$ and its inverse meet that need, and we keep what is proved apart from what is measured.

\subsection{Contributions}
\begin{enumerate}[nosep]
\item A forward involute in binary64 that returned the correctly rounded double at each of 50{,}000 committed test
  angles in $0<|\alpha|\le3\pi/4$ (largest error $0.4999681$ ulp against an 80-digit reference; Section~\ref{sec:results}).
\item An inverse involute that is \emph{proved} faithfully rounded for $10^{-200}\le y<2$, with error at most
  $0.6$ ulp, and a bracket around the root whose containment is proved for $0<y\le10^6$ and was tested with 0 failures at
  47{,}083 points
  (Section~\ref{sec:proofs}).
\item Bit-identical outputs, by hash over the committed samples and an edge-case set, on x86-64 and aarch64 Linux,
  Apple-silicon macOS, AWS Lambda (x86-64 and arm64) and an NVIDIA DGX Spark, in two language runtimes
  (Section~\ref{sec:repro}).
\item A machine-checked proof harness of 32 claims, with mutation tests that plant faults and require each to be caught, and an explicit list of the inputs the proofs trust.
\item A mapping of real gear geometry ($z=5$ to $500$, $\alpha=14.5^\circ$ to $30^\circ$, external and internal) onto the input ranges, and the size, in micrometres, of involute and inverse-involute errors in inspection, manufacturing and CAD
  (Sections~\ref{sec:gears}--\ref{sec:mfg}).
\item A measured comparison of the inverse with public methods (Newton iterations, bisection, explicit
  approximations) on accuracy, time per call and cross-runtime bit identity, over the gear range
  $0.5^\circ$ to $60^\circ$ (Section~\ref{sec:invbench}).
\item A reproducible comparison of naive evaluations in binary16, binary32, binary64 and a 32-bit fixed-point
  format (Table~\ref{tab:precision}).
\end{enumerate}

\subsection{Notation}\label{sec:notation}
Table~\ref{tab:notation} lists the symbols used in the paper. Angles are in radians inside every formula, even where the
text quotes them in degrees. Lengths are in millimetres unless stated; errors in lengths are given in micrometres ($\mu$m). A multiple of the module is written with $m$ directly after the number ($1.25m$, $0.05m_n$); the metre is written with a space (1~m). Subscripts follow ISO~21771-1 and ANSI/AGMA~2002-D19 where they exist.

{\footnotesize
\renewcommand{\arraystretch}{1.1}\setlength{\LTcapwidth}{\textwidth}
\begin{longtable}{@{}>{\raggedright}p{0.22\textwidth}>{\raggedright}p{0.56\textwidth}>{\raggedright\arraybackslash}p{0.14\textwidth}@{}}
\caption{Notation. Where a symbol has two meanings, the Meaning column says which section uses which.}\label{tab:notation}\\
\toprule
Symbol & Meaning & Unit \\
\midrule
\endfirsthead
\toprule
Symbol & Meaning & Unit \\
\midrule
\endhead
\bottomrule
\endfoot
\multicolumn{3}{@{}l}{\emph{The function and its inverse}} \\
$\inv(\alpha)$, $\inv\phi$ & involute function, $\tan\alpha-\alpha$; the polar angle of an involute point from where the curve leaves the base circle & rad \\
$y$ & value of the involute function, $y=\inv\phi$; the input of the inverse & rad \\
$\inv^{-1}(y)$ & inverse involute: the angle in $(0,\pi/2)$ whose involute is $y$ & rad \\
$a$ & a binary64 test angle (Sections~\ref{sec:oracle}--\ref{sec:results} and the tables); elsewhere the centre distance & rad \\
$\delta(\cdot)$ & an error in the quantity that follows, e.g.\ $\delta y$, $\delta s_y$ & as quantity \\
\addlinespace
\multicolumn{3}{@{}l}{\emph{Angles on the gear}} \\
$\alpha$ & nominal (reference) pressure angle of the basic rack; in Sections~\ref{sec:method}--\ref{sec:proofs} also the true root of the inverse & rad ($^\circ$ in text) \\
$\alpha_n$, $\alpha_t$ & normal and transverse pressure angle of a helical gear & rad \\
$\phi$, $\phi(R)$ & local pressure angle at radius $R$, from $\cos\phi=r_b/R$ & rad \\
$\phi_M$, $\alpha_k$ & pressure angle at the centre of the measuring pin or ball (transverse) & rad \\
$\alpha_c$ & pressure angle at the pin's contact point on the flank & rad \\
$\alpha_w$ & working (operating) pressure angle of a gear pair & rad \\
$\beta$ & helix angle at the reference diameter & rad \\
$\psi(R)$ & polar half-angle of the tooth at radius $R$, Eq.~\eqref{eq:halfangle} & rad \\
$t$ & roll angle of the involute ($t=\tan\phi$) & rad \\
\addlinespace
\multicolumn{3}{@{}l}{\emph{Gear size and shape}} \\
$z$, $N$ & number of teeth ($z<0$ for internal gears in ANSI/AGMA~2002-D19; $N$ in Appendix~\ref{app:pins}) & -- \\
$z_0$, $x_0$ & tooth count and shift of a pinion-type cutter & -- \\
$m$, $m_n$ & module; normal module of a helical gear & mm \\
$x$, $x_1$, $x_2$ & profile shift coefficient (per gear of a pair) & -- \\
$r$, $d$ & pitch (reference) radius $r=mz/2$; pitch diameter $d=2r$ & mm \\
$r_b$, $d_b$ & base radius $r_b=r\cos\alpha$; base diameter & mm \\
$R$, $d_y$ & a radius on the flank; a diameter at which thickness is taken & mm \\
$d_a$, $r_{a2}$, $r_{b2}$ & tip diameter; tip and base radius of the mating gear & mm \\
$b$ & face width & mm \\
$s$, $s_n$, $s_y$ & reference tooth thickness; normal thickness; arc thickness at $d_y$ & mm \\
$a$, $a_w$ & standard and working centre distance of a pair & mm \\
$p_b$ & base pitch, $\pi m\cos\alpha$ & mm \\
$L$ & roll length along the line of action, $L=r_bt$; with a subscript (TIF, SAP, EAP, P) the roll length to that point & mm \\
$h'$, $h_{a0}$, $\rho_{a0}$, $\rho_{\rm fP}$ & end of the rack's straight flank; cutter addendum; cutter tip radius; basic-rack tip radius & mm \\
TIF, SAP, LPSTC, HPSTC & true involute form point; start of active profile; lowest and highest points of single tooth contact & -- \\
\addlinespace
\multicolumn{3}{@{}l}{\emph{Inspection}} \\
$d_p$, $D_d$ & diameter of the measuring pin or ball ($D_d$ in ANSI/AGMA~2002-D19) & mm \\
$M$ & measurement over (or between) two pins or balls & mm \\
$d_c$ & diameter of the pin's contact point on the flank & mm \\
$d_{\rm Lf}$, $d_{\rm La}$ & lowest and highest usable involute diameters (root and tip limits) & mm \\
$d_{Cf}$, $d_{Fa}$, $L_\alpha$ & start and end diameters of the ISO~1328-1 profile evaluation; its length along the roll & mm \\
$L_{AE}$ & roll length of the active profile, from the SAP to the tip & mm \\
$\theta_{\rm meas}$, $\theta_{\rm nom}$ & measured and nominal polar angle of a flank point & rad \\
$f$, $e$ & profile deviation along the line of action; a normal form deviation on the contacted flanks & $\mu$m \\
$k$ & odd-tooth factor: 1 for even $z$, $\cos(\pi/2z)$ for odd $z$ & -- \\
gain & change in $M$ per unit normal form deviation on the contacted flanks & -- \\
$F_\alpha$, $f_{f\alpha}$, $f_{H\alpha}$ & total, form and slope profile deviation (ISO~1328-1); with subscript T, the tolerance & $\mu$m \\
$f_{pT}$, $F_{pT}$, $F_{\beta T}$, $f_{f\beta T}$, $f_{H\beta T}$ & ISO~1328-1 pitch and helix tolerances (Appendix~\ref{app:mfg}) & $\mu$m \\
$A$ & ISO~1328-1 tolerance class & -- \\
\addlinespace
\multicolumn{3}{@{}l}{\emph{Numerics}} \\
binary64, binary32, binary16, Q3.28 & IEEE~754 double, single and half precision; a 32-bit fixed-point format with 28 fractional bits & -- \\
$\ulp(v)$ & unit in the last place: the gap between adjacent binary64 numbers at $v$ & -- \\
$\RN(\cdot)$ & rounding to the nearest binary64 number (ties to even) & -- \\
$\ell$, $u$ & lower and upper ends of the enclosure returned for the root & rad \\
EPS\_T, N-TRAJ & a trusted internal error bound used by the inverse proof; the trusted list for Claim~\ref{cl:avail} (Section~\ref{sec:proofs}) & -- \\
\addlinespace
\multicolumn{3}{@{}p{0.95\textwidth}@{}}{\emph{Other}: symbols in the historical appendix ($\omega$, $\theta$, $AB$, $RS$) and the four-bar linkage $O_1M_1M_2O_2$ are defined where used.} \\
\end{longtable}
}

\input{sec_gears}

\section{What the two functions compute}\label{sec:method}
Both functions work in IEEE~754 binary64 with round-to-nearest-even~\cite{ieee754}, written $\RN(\cdot)$. The functions are implemented in JavaScript (ES-module and CommonJS forms) and in Python, and a test requires the two to agree bit for bit. The result does not depend on the platform's math library (such as \code{tan}) or on fused multiply-add; a test gate
fails if either is used. The algorithms are proprietary and are not described here. Readers can check the results
against the test vectors and output hashes (Section~\ref{sec:repro}).

\paragraph{Forward function.} It returns $\inv(\alpha)$ for $0<|\alpha|\le3\pi/4$. The function is odd. At $\alpha=0$ the
input is returned unchanged (so $-0$ stays $-0$); for $|\alpha|>3\pi/4$ or non-finite input the result is NaN.

\paragraph{Inverse function.} For a target $y>0$ it returns the angle $\alpha\in(0,\pi/2)$ with $\inv(\alpha)=y$, which is
unique because $\inv$ is strictly increasing there. The accuracy claims cover $10^{-200}\le y<2$ (proved) and
$y<10^{-200}$ (measured only). All calculator inputs we know of fall well inside $y<2$. Inputs $y\ge2$ are handled, but
the accuracy claims do not cover them. For $y\le0$ the function returns $0$; behaviour for non-finite $y$ is a documented
convention, and we make no general claim about refusing invalid input.

\paragraph{Enclosure.} A separate entry point returns, for $0<y\le10^6$, the solver's value together with two doubles $\ell$ and $u$ with $\ell<\alpha<u$ for the true root $\alpha$. If it cannot return a bracket of 4 ulp or less, it reports an error. Containment is proved and also tested (Section~\ref{sec:proofs}).

\section{What is proved}\label{sec:proofs}
A harness checks the proofs by machine, using \code{mpmath} interval arithmetic~\cite{mpmath,moore2009}, exact
rational arithmetic, and the SMT solvers z3~\cite{z3} and cvc5~\cite{cvc5}. A mutation mode re-runs the harness with one
deliberate fault planted at a time and requires every fault to be caught. The shipped source text is pinned, so any
change to the code invalidates the affected proofs. The approach follows earlier work on certified elementary functions~\cite{dedinechin2011,ziv1991}.

\subsection{The inverse}
\begin{claim}[faithful inverse]\label{cl:inverse}
For every binary64 $y$ with $10^{-200}\le y<2$, the returned value is within $0.6$ ulp of the true root, and
is therefore one of the two doubles adjacent to it (faithful rounding). Below $10^{-200}$ the claim does not apply; that
range is measured only (Section~\ref{sec:results}).
\end{claim}
The proof covers every step from input to returned double, including every rounding. It rests on one assumption, an internal error bound we call EPS\_T. The harness checks EPS\_T, but no one outside the project has derived it, so we list it as a trusted input. An independent derivation, for
example in Gappa~\cite{dedinechin2011} or a proof assistant, is the first item of future work.

\subsection{The enclosure}
\begin{claim}[enclosure always returned, given N-TRAJ]\label{cl:avail}
If the trusted list N-TRAJ below holds, then for every binary64 $y$ with $2\le y\le10^6$ the enclosure entry point returns a
bracket and raises no error.
\end{claim}
A composition checker verifies 8 of 8 obligations for this claim and catches 18 of 18 planted faults. The trusted list
N-TRAJ is: EPS\_T; binary64 round-to-nearest without FMA, as on x86\_64 and in the V8 JavaScript engine; lemmas proved by
z3 and cvc5; interval-arithmetic proofs; and identities replayed numerically but proved by hand.
Claim~\ref{cl:avail} therefore holds only if N-TRAJ holds. It says that a bracket is returned. The next claim says that a returned bracket is correct.
\begin{claim}[returned bracket contains the root]\label{cl:contain}
For every binary64 $y$ with $0<y\le10^6$, a bracket returned by the enclosure entry point satisfies
$\ell<\alpha<u$ for the true root $\alpha$. This claim does not depend on N-TRAJ.
\end{claim}
Eight of the harness's 32 claims concern the bracket, including Claim~\ref{cl:avail}. Together they show that each end of
the bracket is accepted only on a certificate that correctly shows which side of the root it lies on, at every finite
candidate; that the certificates cover every input $0<y\le10^6$ with no gap; that non-finite values are never
certified; and that a bracket wider than 4 ulp is refused rather than returned. Containment is also tested: the test
harness compared returned brackets with the 80-digit reference at 47{,}083 points with $0<y\le10^6$, and every bracket
contained the reference root (0 failures, 0 errors). Only Claim~\ref{cl:avail}, that a bracket is returned rather
than refused for $2\le y\le10^6$, is relative to N-TRAJ; containment is not.

\subsection{What is \emph{not} proved}
We have no proof that the forward function is correctly rounded, or even that its error is below $0.51$ ulp everywhere.
The forward result in this paper is a measurement. For $y\ge2$ the plain inverse is not claimed faithful; there only the bracket is proved (Claims~\ref{cl:avail}
and~\ref{cl:contain}).

\subsection{Harness status}
On commit \code{650722cf} of the development branch (20~November~2026; Python 3.13.5, \code{mpmath} 1.4.1, x86\_64), the composition checker passed 8 of 8 obligations and the harness proved all 32 of its internal claims, with Claim~\ref{cl:avail} (enclosure returned) proved relative to N-TRAJ. The 32 are the harness's own statements; this paper does not number or list them. We reran both on the branch that introduced the 50{,}000-point samples (20~November~2026, same toolchain; solver code and proofs unchanged) and got the same result. For this version we reran both on commit \code{69e27bf8} (23~November~2026; Python 3.12.3, \code{mpmath} 1.4.1,
Node.js 24.22.3): the harness proved all 32 claims, with none unproved or open, and the composition checker passed 8 of 8
obligations. Of the 32, 12 concern the forward function and 21 the inverse and its bracket; one, a step the two share, is
counted in both. The forward-function claims do not prove its accuracy bound, which remains a measurement. The other
Node.js 24.22.3 reruns cover the precision tables (Table~\ref{tab:precision}), the inverse benchmark (Section~\ref{sec:invbench}), and the naive comparison and timing of Section~\ref{sec:repro}; the continuous-integration run of Section~\ref{sec:repro} also hashed both functions' outputs on Node.js 24.22.3.

\section{Testing against an 80-digit oracle}\label{sec:oracle}
The reference is \code{mpmath}~\cite{mpmath} at 80 significant digits. For $|a|<0.2$ it sums 80 terms of the
Maclaurin series of $\tan a-a$; otherwise it evaluates $\tan a-a$ directly. Inverse roots are checked at 120 digits with whichever path the root-finding did not use. Both paths are in one library; no second library, such as Arb, was used. Errors are measured in
ulps of binary64 at the reference value and scored against the \emph{signed} reference, so a wrong sign counts as an error.

The sample sets are committed to the repository as files of binary64 bit patterns, with SHA-256 digests in a
manifest, so that every run, on every machine, scores exactly the same inputs. The forward set has 25{,}000
angles in $(0,3\pi/4]$, each together with its negation, for 50{,}000 angles; the 25{,}000 positive angles are the 9{,}144 positive angles of an earlier 18{,}288-angle test set plus 15{,}856 reproducible pseudo-random angles, drawn alternately uniform in $(0,3\pi/4]$ and log-uniform in $[10^{-12},3\pi/4]$. The inverse set has 50{,}000 points (49{,}999 distinct) with
$10^{-200}\le y<2$: the 44{,}232 points of the earlier test set plus 5{,}768 reproducible pseudo-random points. A separate set of 171
distinct values $y<10^{-200}$ tests the range outside the proof. Every test runs in the Python and both JavaScript
implementations, and they must agree bit for bit (0 JavaScript-versus-Python differences over 50{,}097 forward
inputs and 0 over the 50{,}000 inverse points).

\section{Results}\label{sec:results}

\paragraph{Forward.} The largest measured error over the 50{,}000 angles is $0.4999681$ ulp (at
$a\approx0.0738555$, one of the earlier angles), against an acceptance limit of $0.51$ ulp; 0 of the 50{,}000 are at
or above $0.5$ ulp. Every tested result is therefore correctly rounded. (On the earlier 18{,}288-angle grid the maximum was $0.49997$ ulp.)

\paragraph{Inverse.} The largest measured error over the 50{,}000 points with $10^{-200}\le y<2$ is $0.5555115$ ulp
(at $y\approx0.53359$, one of the pseudo-random points), inside the proved bound of $0.6$ ulp (Claim~\ref{cl:inverse}); 50 points are at
or above $0.5$ ulp. The worst point was rechecked against a root found at 150 digits and gives the same figure.
Over the 44{,}232 earlier points within the committed set the maximum is $0.5226$ ulp.

An earlier draft (17~October~2026) reported $0.5316$ ulp at $y=0.5370916452982755$, over 44{,}232 points. That sample was generated with the platform's math library, so its exact doubles depended on the machine, and the point is not in the committed set. The solver still returns an answer $0.5316$ ulp from the true
root at that $y$ (rechecked for this draft at 150 digits). Committing the sample files removes this source of
variation.

\paragraph{Tiny inputs.} Over the 171 values $y<10^{-200}$, outside the proved range, the
largest measured error is $0.4982$ ulp.

\paragraph{Precision comparison.} Table~\ref{tab:precision} compares naive $\tan a-a$ in several formats with our
forward function at five angles. Each angle is the binary64 value of the label (for $20^\circ$, \code{math.radians(20)}), and the reference is
$\inv$ of that binary64 value. For binary16 and binary32 the angle is first rounded to that format, as a program
storing it there would, and \code{numpy}'s \code{tan} in that format is used. The column ``binary64, CR tan'' uses a
correctly rounded $\tan$ (the 80-digit value rounded once) followed by one binary64 subtraction. Q3.28 is a signed
32-bit fixed-point format with 28 fractional bits; $a$ and the exact $\tan a$ are each rounded to the nearest Q3.28 value (the best case) and then subtracted exactly. Table~\ref{tab:reference} lists the reference
values, and Appendix~\ref{app:ref80} gives them to 80 digits. Table~\ref{tab:inverse} shows the inverse at the
rounded forward values.

\begin{table}[ht]
\centering\small
\input{tables/precision.tex}
\caption{Naive $\tan a-a$ in five arithmetics versus this work, scored against an 80-digit reference.
Generated by the author's script on \tabmachine{} (Python \tabpython, numpy \tabnumpy, mpmath \tabmpmath,
Node \tabnode) against repository commit \code{\tabcommit}, whose solver code is unchanged from \code{650722cf}. Angles are
illustrative points, not the 50{,}000-angle sample.}
\label{tab:precision}
\end{table}

\begin{table}[ht]
\centering\small
\input{tables/reference.tex}
\caption{80-digit reference values (first 30 significant digits shown). The last column compares the 80-digit
value with an independent 120-digit evaluation.}
\label{tab:reference}
\end{table}

\begin{table}[ht]
\centering\small
\input{tables/inverse.tex}
\caption{The inverse at $y=\RN(\inv(\alpha))$ for the angles of Table~\ref{tab:precision}, scored against the
root computed at 100 digits (like the 80-digit reference used elsewhere, far beyond binary64, so the choice does not change the scores).}
\label{tab:inverse}
\end{table}

Three observations follow from Table~\ref{tab:precision}. First, naive binary64 loses digits in proportion to
$\log_{10}(3/a^2)$, as predicted, and the correctly rounded $\tan$ does no better (Section~\ref{sec:hard}). Second, binary16 has no correct digit at small angles, and binary32 keeps between
one and seven. Third, fixed point fails completely at $a=10^{-3}$, where $\inv(a)\approx3.3\times10^{-10}$ is below the
Q3.28 resolution of $2^{-28}\approx3.7\times10^{-9}$, even with an exact $\tan$.

\subsection{The inverse against public methods}\label{sec:invbench}
We compared the inverse with methods common in gear software, all written in plain binary64 JavaScript; every result in this subsection is measured.
\begin{itemize}[nosep]
\item \emph{Newton's method on $\tan\alpha-\alpha-y$}, in three variants. Two use the textbook residual, one seeded
  with $(3y)^{1/3}$ and one with Cheng's economized seed $1.440859\,y^{1/3}-0.3660584\,y$. The third uses Cheng's seed with
  a residual evaluated from the Maclaurin series of $\tan\alpha-\alpha$ below $0.5$~rad, which avoids cancellation.
\item \emph{Explicit approximations:} Cheng's two-term formula~\cite{cheng1992}, and the truncated series reversion
  $\alpha=q-\tfrac{2}{15}q^3+\tfrac{3}{175}q^5-\cdots$ in $q=(3y)^{1/3}$ (the Lam/Polder-type form), with 3, 7 and 12 terms.
\item \emph{Bisection} on $(0,\pi/2)$ with the series residual, run to interval collapse.
\end{itemize}
The test set has 16{,}500 binary64 targets $y$. Of these, 15{,}000 are $y=\RN(\inv(\alpha))$ for $\alpha$ evenly spaced from
$0.05^\circ$ to $75^\circ$, and 1{,}500 are log-spaced from $10^{-15}$ to $10^{-6}$. The reference is the 80-digit
\code{mpmath} root. Errors are in ulps of binary64 at the reference root. Correct digits are $-\log_{10}$ of the relative
error. Times are the median of 51 timed passes over the whole set in a fresh Node.js 24.22.3 process on one x86-64 core,
with three processes per method, in sorted input order. In two checks on Node 24.22.3, shuffling the order changed the times by at most 7\% for this work, the Newton variants and the 7-term series, by up to 12\% for bisection, and by under 2\,ns for Cheng's formula.

\begin{figure}[ht]
\centering
\includegraphics[width=0.85\textwidth]{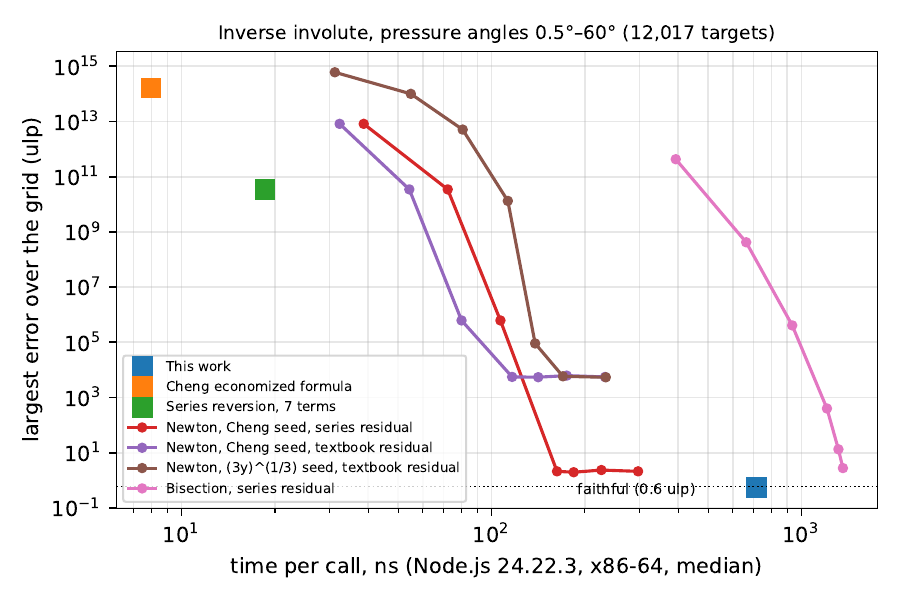}
\caption{Largest error (ulp) against time per call, $0.5^\circ$ to $60^\circ$ (Node.js 24.22.3, x86-64, median). The dotted line
is the 0.6-ulp faithful bound. Only this work falls below it.}
\label{fig:invbench}
\end{figure}

Over the gear range, $0.5^\circ$ to $60^\circ$ (12{,}017 targets), the inverse's largest error is $0.535$ ulp and its mean is
$0.249$ ulp. Of its results, 99.9\% are correctly rounded, every one is faithful, and the worst case keeps 15.9 correct
digits. No public method matched this accuracy at any time budget we tried (Tables~\ref{tab:invbench} and~\ref{tab:invbudget}, Figure~\ref{fig:invbench}). Textbook Newton stalls at about 5{,}600 ulp (12.0 digits) however long it runs; its errors above 100 ulp all lie below $4^\circ$, and from $20^\circ$ to $60^\circ$ it stays within 4.4 ulp. Below $0.5^\circ$ it stalls at $2\times10^{9}$ ulp (6.4 digits), and its usual stopping rule
$|\Delta\alpha|<10^{-15}\alpha$ is never met, so it runs to its cap of 60 iterations. With a cancellation-free residual,
Newton reaches 2.1 ulp (15.3 digits) after four iterations from Cheng's seed and gets no better. Bisection to collapse ends
at 2.8 ulp. The explicit formulas are fast but keep 1.6 (Cheng) to 7.1 (12-term series) digits in this range.

This accuracy costs time. In the gear range the inverse takes 0.72\,\textmu s per call. That is about $1.9\times$ a textbook Newton loop (0.37\,\textmu s) and $4.9\times$ a cancellation-free Newton loop (0.15\,\textmu s), and about half the cost of bisection (about 1.3\,\textmu s). In return it gives the most correct digits at budgets of 800\,ns and above (Table~\ref{tab:invbudget}), and the same bits on every runtime we tried. Tables~\ref{tab:invbench} and~\ref{tab:invbudget} and Figure~\ref{fig:invbench} cover $0.5^\circ$ to $60^\circ$ ($y<0.685$), inside the proved range. The full set runs to $75^\circ$ ($y=2.42$); its 397 targets with $y\ge2$ ($\alpha\ge73.0^\circ$) enter only the bit-identity comparison below. Nothing here extends the faithful-rounding claim beyond $y<2$.

\begin{table}[ht]
\centering\scriptsize
\setlength{\tabcolsep}{3pt}
\begin{tabular}{@{}lrrrrrr@{}}
\toprule
Method & max err. & mean err. & correctly & worst-case & ns & runtimes with \\
 & (ulp) & (ulp) & rounded & digits & per call & differing bits$^{\dagger}$ \\
\midrule
\textbf{This work} & \textbf{0.535} & \textbf{0.249} & \textbf{99.9\%} & \textbf{15.9} & 716 & \textbf{0 of 8} \\
Newton, Cheng seed, series residual & 2.11 & 0.386 & 71.1\% & 15.3 & 146 & 3 of 6 \\
Newton, Cheng seed, textbook residual & $5.6\times10^{3}$ & 45.5 & 37.0\% & 12.0 & 375 & 3 of 6 \\
Newton, $(3y)^{1/3}$ seed, textbook residual & $5.7\times10^{3}$ & 45.7 & 36.6\% & 12.0 & 431 & 3 of 6 \\
Bisection to collapse, series residual & 2.77 & 0.599 & 47.4\% & 15.2 & 1{,}288 & 3 of 6 \\
Series reversion, 12 terms & $3.6\times10^{8}$ & $1.1\times10^{7}$ & 34.6\% & 7.1 & $\approx$24 & not measured \\
Series reversion, 7 terms & $3.4\times10^{10}$ & $2.4\times10^{9}$ & 13.1\% & 5.1 & 19 & not measured \\
Cheng economized formula & $1.6\times10^{14}$ & $2.1\times10^{13}$ & 0\% & 1.6 & 7 to 8 & 1 of 6 \\
\bottomrule
\end{tabular}
\caption{Inverse involute, pressure angles $0.5^\circ$ to $60^\circ$, 12{,}017 binary64 targets, 80-digit reference,
Node.js 24.22.3 on x86-64. Textbook Newton never reaches 1 ulp: it averages 14 to 15 iterations and stops at its cap of 60 on 2{,}131 of the 12{,}017 targets, all below $17^\circ$. Bisection takes about 60 halvings.
$^{\dagger}$Runtimes of Table~\ref{tab:invxrt} that returned at least one different bit on the full 16{,}500-point set,
out of the other runtimes on which the method was run: 8 for this work, 6 (JavaScript only) for the baselines.}
\label{tab:invbench}
\end{table}

\begin{table}[ht]
\centering\small
\begin{tabular}{@{}lrrrrrr@{}}
\toprule
Budget per call & 50\,ns & 100\,ns & 200\,ns & 400\,ns & 800\,ns & 1.6\,\textmu s \\
\midrule
This work & -- & -- & -- & -- & 15.9 & 15.9 \\
Newton, Cheng seed, series residual & 2.8 & 5.1 & 15.4 & 15.4 & 15.4 & 15.4 \\
Newton, Cheng seed, textbook residual & 2.8 & 9.9 & 12.0 & 12.0 & 12.0 & 12.0 \\
Bisection, series residual & -- & -- & -- & 4.1 & 7.1 & 15.2 \\
Cheng economized formula & 1.6 & 1.6 & 1.6 & 1.6 & 1.6 & 1.6 \\
\bottomrule
\end{tabular}
\caption{Worst-case correct digits reached within a time budget, $0.5^\circ$ to $60^\circ$. Each iterative method gets
the largest iteration count that fits the budget. ``--'' means the method cannot return a result within that budget.}
\label{tab:invbudget}
\end{table}

\paragraph{Same answer on every machine.} We ran the inverse on all 16{,}500 targets in nine runtimes: Node.js 18.20.8,
20.20.2, 22.23.5 and 24.22.3 on x86-64; Node.js 22.21.1 and 24.22.3 on arm64; Bun 1.4.4, which uses JavaScriptCore rather
than V8, on x86-64; and the Python implementation on CPython 3.13.5 (x86-64) and 3.13.18 (arm64). Every runtime returned the same bits as the Node.js 24.22.3 x86-64 reference on all 16{,}500 inputs (Table~\ref{tab:invxrt}). The baseline methods, which call the runtime's \code{Math.tan}, returned different bits on some runtimes. Against Node 24 on x86-64, textbook Newton differed on 1{,}556 to 1{,}693 inputs under Bun
and on 269 to 320 under V8 on arm64. The series-residual Newton differed on 766 and 185, and bisection on 338 and 85.
Across V8 versions on the same CPU, the baselines matched. The arm64 runs used \code{qemu-aarch64} user-mode emulation on an
x86-64 host, not native hardware.

\begin{table}[ht]
\centering\scriptsize
\setlength{\tabcolsep}{3pt}
\begin{tabular}{@{}llrrrr@{}}
\toprule
Runtime & Engine, CPU & this work & textbook Newton & series-residual & bisection \\
 & & & (Cheng seed) & Newton & \\
\midrule
Node.js 24.22.3 (reference) & V8, x86-64 & 0 & 0 & 0 & 0 \\
Node.js 22.23.5, 20.20.2, 18.20.8 & V8, x86-64 & 0 & 0 & 0 & 0 \\
Node.js 24.22.3, 22.21.1 & V8, arm64 (qemu) & 0 & 269 & 185 & 85 \\
Bun 1.4.4 & JavaScriptCore, x86-64 & 0 & 1{,}556 & 766 & 338 \\
CPython 3.13.5, 3.13.18 & x86-64, aarch64 (qemu) & 0 & -- & -- & -- \\
\bottomrule
\end{tabular}
\caption{Inputs, out of 16{,}500, on which the output bits differed from Node.js 24.22.3 on x86-64. ``--'' means not run, since the
baselines are JavaScript only.}
\label{tab:invxrt}
\end{table}

\FloatBarrier

\input{sec_practice}

\section{Reproducibility and cross-platform behaviour}\label{sec:repro}
\paragraph{Bit-identical results.} The functions use neither the platform's math library nor fused multiply-add (Section~\ref{sec:method}). ECMAScript specifies binary64
round-to-nearest-even for every Number operation and does not permit contraction~\cite{ecma262}, so any
conforming JavaScript engine should produce identical bits, and CPython inherits the same correctly rounded
basic operations from C.

We checked bit identity in one continuous-integration run (GitHub Actions run 44889953350, 21~November~2026,
on commit \code{927fb121}; later commits change only the checking script, not the solver). Three machines took part:
x86-64 Linux (ubuntu-24.04, AMD EPYC), native aarch64 Linux (ubuntu-24.04-arm), and macOS~15 on a virtual Apple M2 Pro. GitHub's hosted runners are virtual machines, so these are cloud hosts, not bare-metal hardware.
Each ran CPython (3.12.3 on Linux, 3.15.2 on macOS) and Node.js 20.20.2, 22.23.5 and 24.22.3, each in both ES-module and
CommonJS form. For every runtime the run hashed the outputs (SHA-256 of the 64-bit patterns in input order) of the forward
function on the 50{,}000-angle sample, the inverse on the 50{,}000-point sample, and both functions on a separate set of
5{,}180 edge-case inputs. The edge cases cover zero, signed zero, subnormals, negatives, values near $\pi/2$ and beyond
$3\pi/4$, $y\ge2$, $y<10^{-200}$, and $\pm\infty$ and NaN. Every hash matched
the committed reference on all three machines.

The one difference is between the two languages, not between machines: for
$y=+\infty$ and $y=\mathrm{NaN}$ the Python inverse returns $+0$, while the JavaScript inverse returns NaN, so the
edge-case inverse set has one reference hash for CPython and another for Node.js. Each matched on every machine. Every
other input, including $-\infty$ and $y\le0$ (both return $+0$), is bit-identical across the two languages. This non-finite case is outside the paper's accuracy claims.

\paragraph{AWS Lambda.} On 21~November~2026 the same hashes were computed in AWS Lambda on both of its processor
architectures, x86-64 (Intel) and arm64 (AWS Graviton), each with Node.js~22 and Python~3.12. All 16 forward and inverse
output hashes were bit-identical across the two architectures and the two runtimes, and equal to the hashes of the GitHub
Actions platform matrix (run 44941701706). This run checks bit identity only, not accuracy or speed.

\paragraph{NVIDIA DGX Spark.} On 21~November~2026 the platform check was run on an NVIDIA DGX Spark (GB10, aarch64, Linux
6.17, Arm Cortex-X925 and Cortex-A725 cores) on commit \code{0da824cc}, inside a Docker container. It used Python~3.11.2
and Node.js~20.20.2 in both ES-module and CommonJS form. All 12 runtime-and-function hashes matched the committed
reference for the 50{,}000 forward inputs, the 50{,}000 inverse inputs and the 5{,}180 edge-case inputs, and the check
reported PASS. A second run pinned to one Cortex-X925 core also passed; we cannot tell from the check's output which core
type did the work in the first run.

Table~\ref{tab:platforms} summarises the platforms.

\begin{table}[ht]
\centering\footnotesize\setlength{\tabcolsep}{4pt}
\begin{tabular}{@{}llp{0.36\textwidth}c@{}}
\toprule
Platform & CPU & Runtimes & Match \\
\midrule
GitHub, ubuntu-24.04 & x86-64 (AMD EPYC) & CPython 3.12.3; Node 20, 22, 24 (ESM, CJS) & yes \\
GitHub, ubuntu-24.04-arm & aarch64 & CPython 3.12.3; Node 20, 22, 24 (ESM, CJS) & yes \\
GitHub, macOS~15 & Apple M2 Pro (virtual) & CPython 3.15.2; Node 20, 22, 24 (ESM, CJS) & yes \\
AWS Lambda & x86-64 (Intel) & Python 3.12; Node 22 & yes \\
AWS Lambda & arm64 (Graviton) & Python 3.12; Node 22 & yes \\
DGX Spark (Docker) & aarch64 (GB10) & Python 3.11.2; Node 20.20.2 (ESM, CJS) & yes \\
\bottomrule
\end{tabular}
\caption{Platforms on which the output hashes were checked (21~November~2026; GitHub = GitHub Actions run
44889953350). ``Yes'' means every output hash
equalled the reference (for Python and Node.js, the reference of that language; see the non-finite-input note).}
\label{tab:platforms}
\end{table}

{\sloppy Not tested: other operating systems, other processors and other language runtimes. We make no claim for them. The platform check can be rerun on any machine that has the test vectors. It exits with 0 when every hash matches,
with 1 when any output bit differs (and names the input and both results), and with 2 when it cannot run.\par}

For contrast, under Node.js 24.22.3 the naive \code{Math.tan(a) - a} differs between arm64 and x86-64 on 442 of 58{,}392
forward inputs, the same count as in the author's first platform check (20~November~2026). These 58{,}392 inputs are
that check's input set, made before the 50{,}000-angle sample existed: the 9{,}144
positive angles of the earlier 18{,}288-angle test set (Section~\ref{sec:oracle}), the four angles $10^{-3}$, $10^{-2}$
and $10^{-1}$~rad and $20^\circ$, and 20{,}000 reproducible pseudo-random angles uniform in $(0,3\pi/4]$, each with its
negation (58{,}296 angles); 92 out-of-domain values (the 40 doubles just above $3\pi/4$, and 2.4, $\pi$, 4.5, 10,
$10^{10}$ and $10^{300}$, each with its negation); and zero, $\pm\infty$ and NaN. All 442 differing inputs lie in
$0<|a|\le3\pi/4$, between $10.9^\circ$ and $128.1^\circ$ in magnitude. On the 50{,}000-angle sample the same comparison
differs on 274 inputs. In the first check, \code{numpy.tan(a) - a} also differed between aarch64 and x86\_64 builds
(CPython 3.13.18 and 3.13.5, numpy 2.5.5) on 274 of the 58{,}392, by coincidence the same count; that Python comparison
was not rerun. The arm64 runtimes ran under \code{qemu-aarch64} user-mode emulation on an x86-64 host, not on native
hardware.

\paragraph{How to verify.} Every figure in this paper comes from rerunnable scripts: the forward and inverse
gates, the proof harness, the output hashes and the precision tables. Because the solver is proprietary, outside readers verify the results through test vectors (input bit patterns, outputs and 80-digit reference values) and SHA-256 hashes of the outputs, which any implementation can be compared against; both are published in the companion dataset~\cite{involutevectors2026}. The scripts that produced the tables and figures call the proprietary implementation and are not distributed.

\paragraph{Cost.} The forward function is slower than the naive formula. On Node 22 it takes about 244 to 303\,ns per call over
$(0.01,0.3)$, against 22 to 29\,ns for the naive formula, roughly ten times as long. Rerun on Node 24.22.3 (x86-64), it takes 175 to 185\,ns against 13 to 17\,ns. The inverse takes about 0.72\,\textmu s per call over the gear range, $0.5^\circ$ to $60^\circ$ (Node 24.22.3, x86-64; Section~\ref{sec:invbench}).

\section{Limitations}\label{sec:limits}
\begin{itemize}[nosep]
\item The claims concern each function on its own, in binary64. They do not extend automatically to downstream
  quantities (tooth thickness, centre distance, ratings) that combine the functions with other arithmetic.
\item The forward bound is measured, not proved.
\item The inverse proof covers $10^{-200}\le y<2$; below $10^{-200}$ accuracy is measured only. For $y\ge2$ the plain inverse is not claimed faithful. A returned bracket is proved to contain the root for $0<y\le10^6$, independently of N-TRAJ, and was tested there; only the guarantee that a bracket is returned for $2\le y\le10^6$ is relative to N-TRAJ.
\item EPS\_T is an assumption with no independent derivation, and the faithful-inverse proof uses it.
\item The reference is one arbitrary-precision library with two evaluation paths.
\item Bit-identity is verified on the machines of Table~\ref{tab:platforms} (x86-64 and aarch64 Linux, Apple-silicon
  macOS, AWS Lambda x86-64 and arm64, NVIDIA DGX Spark) and the runtimes named in Section~\ref{sec:repro}, on the committed samples and edge cases, not on every input. The Python and
  JavaScript inverses differ for $+\infty$ and NaN input.
\item The gear envelope of Section~\ref{sec:gears} shows where real inputs fall; it is not a claim about the accuracy of
  tooth geometry, inspection results or ratings computed with these functions, and it covers spur-gear transverse
  geometry with the ISO~53~A rack only.
\item We make no claim about rejecting invalid inputs.
\end{itemize}

\section{Future work}
Future work: an independent derivation of EPS\_T; a proof of the forward bound; an independent second oracle (e.g.\ Arb or MPFR);
bit-identity runs on further hardware, operating systems and runtimes; aligning the two implementations on
non-finite input; and extending the faithful-inverse proof beyond $y<2$.

\section{Conclusions}\label{sec:conclusions}
\begin{itemize}[nosep]
\item \emph{CAD.} A tool without input checks draws an impossible gear and exports a file with every flank missing
  (Figure~\ref{fig:freecad}; the example tooth is already pointed at $x\approx1.22$, yet $x=3.00$ is accepted). Circular-arc
  teeth are not involutes: the measured SolidWorks Toolbox flank departs from the involute over a span of \tbSpan~$\mu$m between
  the SAP and the tip, with a slope deviation just outside class~11 (Section~\ref{sec:mfg}). Binary16 arithmetic costs \fwSixteenFive~$\mu$m
  in tooth thickness on a 1~m gear, and $\tan\phi-\phi$ computed from angles rounded in degrees loses accuracy
  (Appendix~\ref{app:flank}).
\item \emph{Manufacturing.} The cutter, not the arithmetic, sets a generated profile: one arc-minute of hob pressure-angle
  error tilts the worked example's profile by \hobFhTwo~$\mu$m and the $m=10$~mm, $z=100$ gear's by \hobFhTen~$\mu$m, more
  than its class-\hobExcOneTen{} tolerance. A part made from a SolidWorks Toolbox model carries its built-in backlash: each tooth is
  \tbThin~mm thinner at the pitch circle (3.5 times the \tbBL~mm backlash entry).
\item \emph{Inspection.} Where the pin touches matters most: the form-error gain is \exLowGain{} at the TIF against
  \exMidGain{} mid-flank (\exLowFf{} against \exMidFf~$\mu$m for a class-5 form error), and up to \pinLowGainMax{} over the
  envelope. Shortcut inverse formulas cost \pinChengMin{} to \pinChengMax~$\mu$m over pins and \cdCheng~$\mu$m in the centre
  distance of a shifted pair.
\end{itemize}
Done correctly, the involute arithmetic is not the hazard; set-up, input checks and the choice of measurement are.
The forward function returned the correctly rounded double at all 50{,}000 tested angles (measured, not proved), the
inverse is proved faithfully rounded for $10^{-200}\le y<2$, which covers every gear in our envelope, a returned bracket
is proved to contain the root for $0<y\le10^6$ and did so at all 47{,}083 test points, and both functions gave the
same bits on every machine and runtime we tried (Table~\ref{tab:platforms}). 

Software should warn or refuse rather than draw an impossible gear (Section~\ref{sec:inputchecks},
Figure~\ref{fig:freecad}). Naive binary64 puts at most \fwSixtyFourMax~$\mu$m into tooth thickness in our cases, and
\dxPolyOne{} polyline points hold the worked example's flank to 1~$\mu$m. Over pins and in centre distance a converged
binary64 inverse contributes less than $10^{-10}$~$\mu$m, but a shortcut formula's angle error is multiplied by
$\partial M/\partial\phi_M$, which grows with gear size, so shortcuts do the most harm on large gears. The standard
mid-flank pin sizes avoid the gain of up to \pinLowGainMax{} at a low contact.

\section*{Data availability}
The test vectors (input bit patterns, outputs and 80-digit reference values) and the SHA-256 output hashes are published as a companion dataset under CC BY 4.0~\cite{involutevectors2026}.

\section*{Use of AI assistance}
The author is the sole author and takes full responsibility for the content. The problem framing, the
selection and curation of sources (including the historical references), the acceptance criteria (what counts
as proved, what counts as measured, and the limits each gate enforces) and the final verification of every claim
are the author's. AI coding and research assistants were used to implement the code, search the literature, draft
proof scripts and text, and run the test and proof harnesses. No claim in this paper rests on an assistant's
assertion. Every numerical claim is backed by a rerunnable test or a machine-checked proof, and every historical statement was checked against its cited source; the planted-fault (mutation) runs show that the checks fail when the code or the arguments are wrong.

\bibliographystyle{plainnat}
\begingroup\raggedright
\bibliography{refs}
\endgroup

\appendix
\input{app_history}
\input{app_flank}
\input{app_mfg}
\input{app_pins}

\section{Gear-practice methods}\label{app:gears}
All numbers in Sections~\ref{sec:gears}--\ref{sec:mfg} come from the author's Python scripts (geometry in binary64, error studies at 60 to 80 digits with \texttt{mpmath}, this work's values from its JavaScript implementation). The scripts are not distributed, because they call the proprietary implementation.
\begin{itemize}[nosep]
\item \emph{Rack generation.} We simulate the ISO~53~A rack (straight flank, $1.25m$ dedendum, tip fillet $\rho_{\rm fP}$ by the rule of Section~\ref{sec:gears}: $0.38m$ up to $23.16^\circ$, then the largest that fits) rolling
  on the pitch circle. For each radius we take the innermost swept point, and the involute is taken to start where the
  envelope first agrees with~\eqref{eq:halfangle} to within $0.001m$. The analytic TIF uses the end of the rack's straight
  flank, $h_{FfP}=1.25m-\rho_{\rm fP}(1-\sin\alpha)$ below the datum line.
\item \emph{Inverse methods.} Naive Newton starts at $(3y)^{1/3}$ and iterates $\phi\leftarrow\phi-(\tan\phi-\phi-y)/\tan^2\phi$
  in the stated format until the step is below one unit of relative precision. This baseline differs from the textbook Newton of Section~\ref{sec:invbench} in its stopping rule. At the pin angles of Table~\ref{tab:pins} (above $20^\circ$) both are accurate to a few ulp; the 12-digit floor reported in Section~\ref{sec:invbench} occurs only at small angles. Cheng's approximation is
  $1.440859\,y^{1/3}-0.3660584\,y$. Errors are measured against roots found at 80 digits.
\item \emph{Circular arcs.} The best single arc minimises the maximum normal deviation (Nelder--Mead on centre and radius).
  The four-bar ratio is obtained by solving the linkage $O_1M_1M_2O_2$ numerically.
\item \emph{Discretisation.} Greedy maximal segments, with sag measured against 200{,}001 points of the exact flank.
\end{itemize}

\begin{table}[ht]
\centering\small
\input{tables/gear_forward.tex}
\caption{Error in tooth thickness at diameter $d_y$, $\delta s_y=d_y\,\delta(\inv\phi)$, from naive forward evaluation
in three formats ($\alpha=20^\circ$).}
\label{tab:gearfwd}
\end{table}

\section{Reference values to 80 significant digits}\label{app:ref80}
\enlargethispage{3\baselineskip}
Each value is $\inv(a)$ at the binary64 angle $a$ named, from \code{mpmath} at 80 digits (computed with 40 guard
digits). It agrees with a 120-digit evaluation to at least 79 digits.
{\footnotesize
\begin{description}[style=nextline,font=\normalsize\normalfont,leftmargin=1em,nosep]
\input{tables/reference80.tex}
\end{description}}

\end{document}

%% file: tables/meta.tex
\newcommand{\tabcommit}{a28f989b}\newcommand{\tabpython}{3.13.5}\newcommand{\tabnumpy}{2.5.5}\newcommand{\tabmpmath}{1.4.1}\newcommand{\tabnode}{24.22.3}\newcommand{\tabmachine}{x86\_64}

%% file: tables/gear_macros.tex
\newcommand{\envN}{15{,}872}
\newcommand{\envTipPhiMin}{15.36}
\newcommand{\envTipPhiMax}{51.79}
\newcommand{\envTipYMin}{0.00661}
\newcommand{\envTipYMax}{0.3663}
\newcommand{\envTifPhiMin}{0.034}
\newcommand{\envTifPhiMax}{29.52}
\newcommand{\envTifYMin}{\ensuremath{7.2\times10^{-11}}}
\newcommand{\envTifYMax}{0.0510}
\newcommand{\envUndercut}{373}
\newcommand{\envShiftPointed}{104}

\newcommand{\envYShiftMax}{0.5819}
\newcommand{\zminA}{31}
\newcommand{\zminB}{18}
\newcommand{\zminC}{12}
\newcommand{\zminD}{10}

\newcommand{\yOneNm}{\ensuremath{8.4\times10^{-14}}}
\newcommand{\pointedPhiMax}{53.04}

\newcommand{\uStartR}{2.5719}
\newcommand{\uStartPhi}{19.77}

\newcommand{\uXmin}{0.808}
\newcommand{\uGtif}{-3.229}
\newcommand{\uShiftTip}{-0.819}

\newcommand{\exEps}{1.557}
\newcommand{\exPb}{5.9043}
\newcommand{\exTifNum}{37.595}
\newcommand{\arcTbMin}{-40.7}
\newcommand{\arcTbMax}{25.9}
\newcommand{\arcBest}{9.3}
\newcommand{\arcFlat}{17.0}
\newcommand{\arcTbMinB}{-32.9}
\newcommand{\arcTbMaxB}{85.9}
\newcommand{\arcBestB}{12.7}
\newcommand{\arcGamma}{23.54}

\newcommand{\arcLoB}{22.585}
\newcommand{\arcLo}{37.855}
\newcommand{\dxLen}{3.453}
\newcommand{\dxPolyOne}{16}
\newcommand{\dxPolyTen}{6}

\newcommand{\dxArcOne}{3}
\newcommand{\dxArcTenth}{7}
\newcommand{\dxBigLen}{20.48}
\newcommand{\dxBigPolyOne}{19}
\newcommand{\dxBigArcOne}{2}
\newcommand{\stlSpacingSmall}{0.0019}
\newcommand{\stlSpacingBig}{0.031}
\newcommand{\hobPbTwo}{0.63}
\newcommand{\hobFhTwo}{0.97}
\newcommand{\hobLaTwo}{9.19}
\newcommand{\hobIsoTwo}{4.8}
\newcommand{\hobPbTen}{3.13}
\newcommand{\hobFhTen}{5.79}
\newcommand{\hobIsoTen}{9.0}
\newcommand{\cdAw}{504.8296}

\newcommand{\cdCheng}{24.6}
\newcommand{\cdThirtyTwo}{0.017}
\newcommand{\cdSixtyFour}{\ensuremath{3.0\times10^{-11}}}
\newcommand{\pinChengMax}{42.5}
\newcommand{\pinChengMin}{2.5}
\newcommand{\pinThirtyTwoMax}{0.023}
\newcommand{\pinSixtyFourMax}{\ensuremath{4.9\times10^{-11}}}
\newcommand{\dlWorstSixtyFour}{7}
\newcommand{\dlWorstThirtyTwo}{0}
\newcommand{\dlTifSixtyFour}{13}
\newcommand{\dlTifThirtyTwo}{4}

\newcommand{\fwSixtyFourMax}{\ensuremath{6.3\times10^{-12}}}
\newcommand{\sapMinPhi}{0.081}
\newcommand{\sapMinZ}{21}
\newcommand{\sapMinA}{15}

\newcommand{\sapNoValid}{224}
\newcommand{\sapCleanPhi}{3.44}
\newcommand{\sapCleanZ}{31}
\newcommand{\sapCleanA}{14.5}
\newcommand{\sapCleanY}{\ensuremath{7.2\times10^{-5}}}
\newcommand{\thBase}{3.512}
\newcommand{\thSap}{3.516}
\newcommand{\thPitch}{3.142}
\newcommand{\thTip}{1.390}
\newcommand{\hPrime}{2.000}
\newcommand{\lTif}{0.9930}
\newcommand{\lSap}{2.2444}
\newcommand{\zClassA}{31.9}
\newcommand{\zClassB}{17.1}
\newcommand{\hPrimeA}{0.965}
\newcommand{\arcRange}{66.6}
\newcommand{\arcRangeB}{118.8}
\newcommand{\arcBand}{18.5}
\newcommand{\arcBandB}{25.4}
\newcommand{\isoFaFiveA}{8}
\newcommand{\isoFaElevenA}{62}

\newcommand{\arcBestFitA}{8}
\newcommand{\arcBestFitB}{9}
\newcommand{\isoFaBestA}{22}

\newcommand{\chengExcMin}{2}
\newcommand{\chengExcMax}{9}
\newcommand{\cdChengExc}{7}
\newcommand{\fwSixteenExc}{6}
\newcommand{\fwSixteenFive}{20.7}

\newcommand{\hobExcFiveTwo}{5}
\newcommand{\hobExcOneTen}{3}
\newcommand{\hobFhOneTwoIso}{1.2}
\newcommand{\isoFfFiveA}{6}

\newcommand{\pinGears}{15{,}499}
\newcommand{\pinExcl}{373}
\newcommand{\pinLowAkMin}{2.7}
\newcommand{\pinLowAkMax}{29.5}
\newcommand{\pinLowYMin}{\ensuremath{3\times10^{-5}}}

\newcommand{\pinLowGainMax}{42.6}
\newcommand{\pinLowGainMin}{4.1}
\newcommand{\pinProjAkMin}{14.7}

\newcommand{\pinProjGainMax}{7.9}
\newcommand{\pinMidGain}{3.0}
\newcommand{\pinMidGainMax}{7.9}
\newcommand{\pinHighGainMin}{2.3}
\newcommand{\pinHighGainMax}{7.5}
\newcommand{\pinHighYMax}{0.680}
\newcommand{\pinHighAkMax}{59.9}
\newcommand{\pinLowRootPct}{47}
\newcommand{\pinLowProjPct}{0}
\newcommand{\pinDrcLow}{3.3}

\newcommand{\pinDrcHigh}{1.8}
\newcommand{\pinConvMax}{\ensuremath{6\times10^{-11}}}
\newcommand{\pinBtwoMax}{0.047}
\newcommand{\pinChengLowMax}{136}
\newcommand{\pinChengMidMax}{146}
\newcommand{\pinChengHighMax}{777}
\newcommand{\pinChengHighZ}{10}
\newcommand{\pinChengHighA}{30}
\newcommand{\pinNErr}{1{,}933}
\newcommand{\pinKoddMin}{0.951}
\newcommand{\pinStdBelowPct}{9}
\newcommand{\exLowGain}{17.2}
\newcommand{\exMidGain}{4.3}
\newcommand{\exLowFf}{103}
\newcommand{\exMidFf}{26}

\newcommand{\exLowCheng}{0.42}
\newcommand{\exHighCheng}{15}

\newcommand{\intGrid}{15{,}712}
\newcommand{\intGears}{14{,}719}

\newcommand{\intNminLo}{30}
\newcommand{\intNminHi}{72}
\newcommand{\intNminA}{68}
\newcommand{\intNminB}{37}
\newcommand{\intNminC}{32}
\newcommand{\intNminD}{72}
\newcommand{\intZoA}{50}
\newcommand{\intZoB}{26}
\newcommand{\intZoD}{67}

\newcommand{\intDzLo}{5}
\newcommand{\intDzHi}{18}

\newcommand{\intTipAkMin}{3.4}
\newcommand{\intTipAkMax}{29.3}

\newcommand{\intTipYMin}{\ensuremath{7\times10^{-5}}}
\newcommand{\intTipYMax}{0.050}

\newcommand{\intTipGainMax}{33.9}
\newcommand{\intTipDrc}{12.6}
\newcommand{\intTipProjPct}{100}

\newcommand{\intTipCheng}{132}

\newcommand{\intMidGainMax}{8.8}
\newcommand{\intMidDrc}{2.3}
\newcommand{\intMidProjPct}{99.98}

\newcommand{\intMidCheng}{142}

\newcommand{\intRootAkMax}{34.9}

\newcommand{\intRootYMax}{0.089}

\newcommand{\intRootGainMax}{7.7}

\newcommand{\intRootProjPct}{0}

\newcommand{\intRootCheng}{152}

\newcommand{\intTipGainZ}{63}
\newcommand{\intTipGainA}{15}
\newcommand{\intChengZ}{500}
\newcommand{\intChengA}{30}
\newcommand{\intTipLagMin}{0.23}
\newcommand{\intTipLagMax}{4.7}
\newcommand{\intRootTouchPct}{2.2}

\newcommand{\intPbTipsidePct}{40}
\newcommand{\intConvMax}{\ensuremath{6\times10^{-11}}}
\newcommand{\intBtwoMax}{0.047}
\newcommand{\intNErr}{1{,}795}
\newcommand{\extLowYMin}{\ensuremath{3\times10^{-5}}}
\newcommand{\extLowAkMin}{2.7}

\newcommand{\exiTipFf}{58}
\newcommand{\exiMidFf}{38}

\newcommand{\tbR}{8.000}

\newcommand{\tbRcErr}{$2\times10^{-9}$}
\newcommand{\tbDa}{44.000}
\newcommand{\tbDf}{34.998}

\newcommand{\tbRot}{0.337}
\newcommand{\tbOff}{19.374}
\newcommand{\tbOffModel}{19.037}
\newcommand{\tbSpitch}{2.906}
\newcommand{\tbThin}{0.235}
\newcommand{\tbNormOff}{0.110}

\newcommand{\tbDevMin}{67.6}
\newcommand{\tbDevMax}{133.8}
\newcommand{\tbSpan}{66.1}
\newcommand{\tbFa}{57.5}
\newcommand{\tbFf}{23.8}
\newcommand{\tbFh}{39.6}
\newcommand{\tbLa}{8.73}
\newcommand{\tbFaCls}{11}
\newcommand{\tbFfCls}{9}
\newcommand{\tbFaTol}{62}
\newcommand{\tbFfTol}{24}
\newcommand{\tbFhElevenTol}{39}

\newcommand{\tbModelFh}{38.8}
\newcommand{\tbModelFhCls}{11}
\newcommand{\tbRatioMin}{0.9902}
\newcommand{\tbRatioMax}{1.0099}
\newcommand{\tbHalfNom}{1.5692}
\newcommand{\tbHalfCT}{1.452}
\newcommand{\tbBL}{0.067}
\newcommand{\tbChordFull}{2.904}

%% file: sec_gears.tex
\section{Gear geometry and the input envelope}\label{sec:gears}

\subsection{The flank of a spur gear tooth}\label{sec:flank}
\begin{figure}[ht]
\centering
\includegraphics[width=0.82\textwidth]{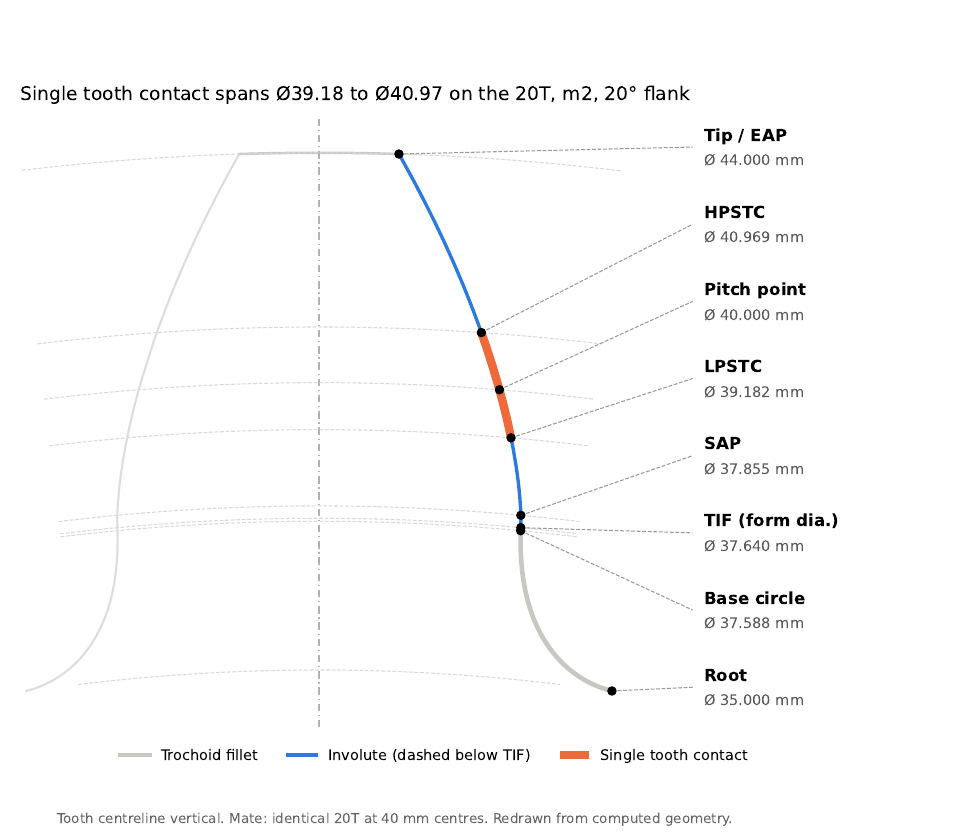}
\caption{Right flank of the worked example ($z=20$, $m=2$~mm, $\alpha=20^\circ$, identical mate at 40~mm centres), redrawn
to the labelling of the author's flank drawing from geometry computed by the author's script. Blue: involute
(dashed below the TIF). Orange: single tooth contact (LPSTC to HPSTC). Grey: trochoid fillet from the simulated rack
generation. Diameters from Table~\ref{tab:example}.}
\label{fig:flank}
\end{figure}

A spur-gear flank is fixed by the tooth count $z$, the module $m$, the nominal pressure angle $\alpha$ and the profile
shift $x$. With pitch radius $r=mz/2$ and base radius $r_b=r\cos\alpha$, the local pressure angle at radius $R$ satisfies
$\cos\phi=r_b/R$, and the polar half-angle of the tooth there is
\begin{equation}\label{eq:halfangle}
\psi(R)=\frac{\pi}{2z}+\frac{2x\tan\alpha}{z}+\inv\alpha-\inv\phi(R).
\end{equation}
The flank point is $(R\cos\psi,\pm R\sin\psi)$ and the arc tooth thickness is $s_y=2R\,\psi(R)$. Every flank point, whether in a drawing, a DXF or STEP file or an inspection nominal, is therefore computed through $\inv\phi$, and recovering $\phi$
from $\inv\phi$ needs the inverse. Module only scales the tooth: $\phi$ and $\inv\phi$ at each named point depend on
$z$, $\alpha$ and $x$ alone. Figure~\ref{fig:flank} shows the named points of the author's worked example ($z=20$,
$m=2$~mm, $\alpha=20^\circ$, ISO~53~A rack, identical mate at 40~mm centres). Appendix~\ref{app:flank} gives the author's
background notes on generation, the formulas for each named point, and the worked example in full
(Table~\ref{tab:example}). Because $\inv\phi\approx\phi^3/3$, the polar angle changes very little low on the flank: in the
example the tooth thickness changes only from \thBase{} to \thSap~mm between the base circle and the SAP.

\subsection{\texorpdfstring{The envelope: $z=5$ to $500$, $\alpha=14.5^\circ$ to $30^\circ$}{The envelope: z = 5 to 500, alpha = 14.5 to 30 degrees}}
We evaluated every integer $z$ from 5 to 500 and every $\alpha$ from $14.5^\circ$ to $30^\circ$ in steps of $0.5^\circ$
(\envN{} gears), with standard full-depth proportions (addendum $1m$, dedendum $1.25m$) and $x=0$. The rack's tip-corner radius is ISO~53's $\rho_{\rm fP}=0.38m$ where it fits. Above $23.16^\circ$ the two corner arcs no longer fit on the $1.25m$-deep rack tooth (tip width $\pi m/2-2.5m\tan\alpha$), so we use the largest radius that does, $\rho_{\rm fP}=\min\bigl(0.38m,\ (\pi m/2-2.5m\tan\alpha)\cos\alpha/(2(1-\sin\alpha))\bigr)$, with a floor of $0.05m$, which never applies in this envelope (it would only above $31.2^\circ$). This gives $0.32m$ at $25^\circ$ and $0.11m$ at $30^\circ$. Because module only scales the tooth (Section~\ref{sec:flank}), these results hold for every module.
\begin{itemize}[nosep]
\item \emph{Tip:} $\phi$ from \envTipPhiMin$^\circ$ to \envTipPhiMax$^\circ$, $y=\inv\phi$ from \envTipYMin{} to \envTipYMax.
\item \emph{Lowest involute point (TIF), gears without undercut:} $\phi$ from \envTifPhiMin$^\circ$ to
  \envTifPhiMax$^\circ$, $y$ from \envTifYMin{} to \envTifYMax.
\item \emph{Lowest working SAP} (identical mate, valid SAP above the TIF, no undercut): $\phi=\sapCleanPhi^\circ$
  ($y=\sapCleanY$, $z=\sapCleanZ$, $\alpha=\sapCleanA^\circ$), close to the author's earlier estimate of about $3^\circ$.
\item \emph{Undercut:} with $x=0$, \envUndercut{} of the gears are undercut ($z$ below \zminA{} at $14.5^\circ$, below \zminB{} at $20^\circ$). The case $z=5$, $\alpha=14.5^\circ$ undercuts heavily: the involute survives only above $R=\uStartR\times m$ (module)
  ($\phi=\uStartPhi^\circ$, above the pitch circle) and everything below is a non-conjugate trochoid. Positive shift is
  the usual remedy, but at the shift needed here ($x=\uXmin$) the tip is already pointed, so the outside diameter must
  also be reduced. Shifted or pointed teeth raise the largest $y$ to at most \envYShiftMax{}, a bound taken as the larger of the tip-circle and pointed-tip values; the flank itself reaches at most $y=0.4995$ ($\phi=55.85^\circ$, Appendix~\ref{app:flank}).
\item \emph{Near the base circle:} $y\to0$ as $\phi\to0$. Even so, on a gear with a 1~m base diameter a point 1~nm above the base circle has $y=\yOneNm$; smaller gears give larger $y$ at the same height.
\end{itemize}
Every point on the involute of every gear in this envelope, shifted or not, has $0<\phi<\pi/2$ and, apart from points
within 1~nm of the base circle, $\yOneNm\le y\le\envYShiftMax$. \emph{The gears in this envelope therefore produce inputs
inside the tested forward range $(0,3\pi/4]$ and inside the proved inverse range $10^{-200}\le y<2$}, by a wide margin
at both ends.  Appendix~\ref{app:flank} gives the undercut and SAP details.

\subsection{Where the numerical problem lives}
\begin{figure}[ht]
\centering
\includegraphics[width=0.9\textwidth]{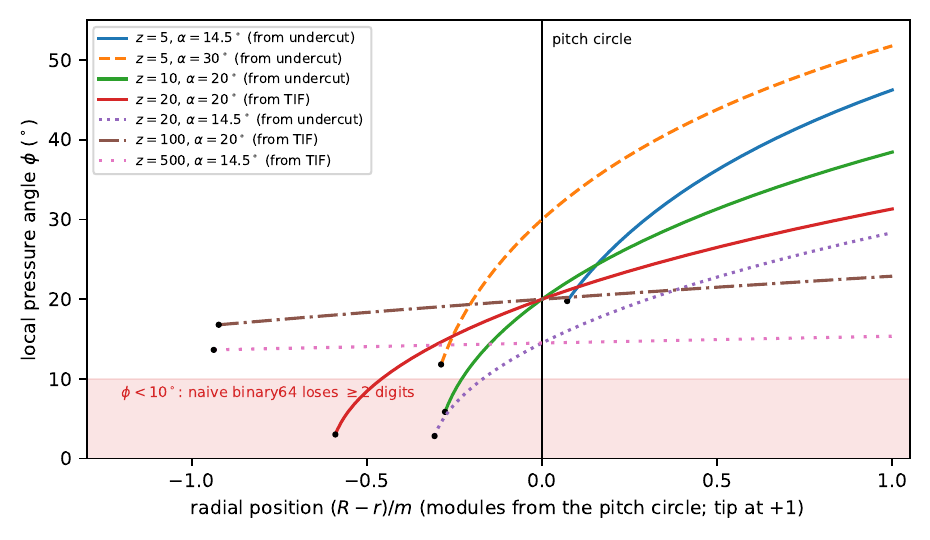}
\caption{Local pressure angle along the involute from its lowest point (dot: TIF, or the start of the involute
left by undercut) to the tip, for gears across the envelope ($x=0$). The shaded band $\phi<10^\circ$ is where
naive binary64 loses at least two significant digits; it lies below the pitch circle.}
\label{fig:phiroll}
\end{figure}

The cancellation in $\tan\phi-\phi$ depends on the local pressure angle, not the nominal one. The local pressure
angle is smallest low on the flank: near the base circle, at the TIF and the SAP, usually below the pitch diameter
(Figure~\ref{fig:phiroll}). The relative error of the naive formula is amplified by $\tan\phi/\inv\phi\approx3/\phi^2$.
Table~\ref{tab:digits} scores the naive formula at representative local pressure angles, as Table~\ref{tab:precision} does. Naive
binary64 keeps \dlTifSixtyFour{} digits at the TIF of the worked example and \dlWorstSixtyFour{} digits 1~nm above the
base circle of a 1~m gear. Naive binary32 keeps \dlTifThirtyTwo{} and \dlWorstThirtyTwo{} digits respectively.
Section~\ref{sec:microns} translates these digits into lengths.

\begin{table}[ht]
\centering\small
\input{tables/gear_digits.tex}
\caption{Correct significant digits of naive $\tan\phi-\phi$ at representative local pressure angles (scoring as
in Table~\ref{tab:precision}). This work's \texttt{inv} keeps 15 to 17 digits at every row.}
\label{tab:digits}
\end{table}

%% file: tables/gear_digits.tex
\begin{tabular}{lrrrrr}
\toprule
location & $\phi$ ($^\circ$) & $y=\inv\phi$ & $\tan\phi/\inv\phi$ & \multicolumn{2}{c}{naive digits: b64 / b32}\\
\midrule
1 nm above base, $d_b=1$ m & 0.003624 & $8.43\times10^{-14}$ & $7.5\times10^{8}$ & 7 & 0 \\
1 $\mu$m above base, $d_b=1$ m & 0.1146 & $2.67\times10^{-9}$ & $7.5\times10^{5}$ & 10 & 1 \\
$z=18$, $20^\circ$, TIF & 1.046 & $2.03\times10^{-6}$ & 8994 & 12 & 3 \\
$z=5$, $14.5^\circ$, involute start & 19.77 & 0.0144 & 25 & 14 & 6 \\
$z=20$, $20^\circ$, TIF & 3.024 & $4.91\times10^{-5}$ & 1076 & 13 & 4 \\
$z=20$, $20^\circ$, SAP & 6.81 & $5.63\times10^{-4}$ & 212 & 14 & 5 \\
$z=20$, $20^\circ$, pitch & 20 & 0.0149 & 24 & 15 & 6 \\
$z=20$, $20^\circ$, tip & 31.32 & 0.0619 & 9.8 & 15 & 6 \\
\bottomrule
\end{tabular}

%% file: tables/precision.tex
\begin{tabular}{lrrrrrr}
\toprule
$\alpha$ (rad) & binary16 & binary32 & binary64 & binary64, CR tan & Q3.28 fixed & this work \\
\midrule
\multicolumn{7}{l}{\emph{Correct significant digits} ($\lfloor -\log_{10}$ rel.\ error$\rfloor$, clipped at 0)} \\
$10^{-3}$ & 0 & 1 & 9 & 9 & 0 & 16 \\
$10^{-2}$ & 0 & 3 & 11 & 11 & 2 & 16 \\
$10^{-1}$ & 1 & 5 & 14 & 14 & 5 & 16 \\
$20^\circ$ (0.349) & 3 & 6 & 15 & 15 & 7 & 16 \\
$1.0$ & 3 & 7 & 15 & 15 & 8 & 16 \\
\midrule
\multicolumn{7}{l}{\emph{Error in binary64 ulps of the reference}} \\
$10^{-3}$ & $6.4\times10^{15}$ & $3.1\times10^{14}$ & $1.3\times10^{6}$ & $1.3\times10^{6}$ & $6.6\times10^{16}$ & 0.077 \\
$10^{-2}$ & $6.3\times10^{15}$ & $1.3\times10^{12}$ & $1.2\times10^{4}$ & $1.2\times10^{4}$ & $3.4\times10^{13}$ & 0.126 \\
$10^{-1}$ & $5.4\times10^{14}$ & $1.0\times10^{10}$ & $2.8\times10^{1}$ & $2.8\times10^{1}$ & $5.9\times10^{10}$ & 0.265 \\
$20^\circ$ (0.349) & $6.8\times10^{12}$ & $2.3\times10^{9}$ & 1.492 & 1.492 & $1.7\times10^{8}$ & 0.492 \\
$1.0$ & $1.9\times10^{12}$ & $1.1\times10^{8}$ & 0.557 & 0.557 & $8.5\times10^{6}$ & 0.443 \\
\bottomrule
\end{tabular}

%% file: tables/reference.tex
\begin{tabular}{lll}
\toprule
$\alpha$ & $\mathrm{inv}(\alpha)$, 80-digit reference (first 30 digits) & agrees with 120-digit run to \\
\midrule
$10^{-3}$ & \texttt{3.33333466666720655759200007552e-10} & 79+ digits \\
$10^{-2}$ & \texttt{3.33346667206371097542171549283e-7} & 79+ digits \\
$10^{-1}$ & \texttt{3.34672085450545113963379422232e-4} & 79+ digits \\
$20^\circ$ (0.349) & \texttt{1.49043838673364433466175910699e-2} & 79+ digits \\
$1.0$ & \texttt{5.57407724654902230506974807458e-1} & 79+ digits \\
\bottomrule
\end{tabular}

%% file: tables/inverse.tex
\begin{tabular}{lll}
\toprule
$y=\mathrm{RN}(\mathrm{inv}(\alpha))$ for $\alpha$ & this work, $\inv^{-1}(y)$ & error (ulp) \\
\midrule
$10^{-3}$ & \texttt{0.001} & 0.0184 \\
$10^{-2}$ & \texttt{0.01} & 0.0385 \\
$10^{-1}$ & \texttt{0.1} & 0.1029 \\
$20^\circ$ (0.349) & \texttt{0.3490658503988659} & 0.1160 \\
$1.0$ & \texttt{1.0} & 0.0913 \\
\bottomrule
\end{tabular}

%% file: sec_practice.tex
\section{Where micrometres count: inspection}\label{sec:microns}

The largest effect on the measurement over pins $M$ is where the pin touches the flank. A pin or ball chosen for convenience can touch low on the flank, near the TIF and below the pitch circle, where form deviation is multiplied most strongly into $M$. On the $z=20$ worked example the gain
is \exLowGain{} at the TIF against \exMidGain{} at mid-flank, so a local form deviation equal to the ISO~1328-1 class-5
tolerance ($f_{f\alpha T}=\isoFfFiveA$~$\mu$m) moves $M$ by about \exLowFf~$\mu$m instead of \exMidFf~$\mu$m. Over all
\pinGears{} external gears we studied, the gain at a low contact reaches \pinLowGainMax{} (Section~\ref{sec:pinsloc},
Table~\ref{tab:pinsloc}). The gain comes from the gear geometry and the measurement set-up, and it exceeds every arithmetic error of a converged inverse. Practical guidance from the
same study and from ANSI/AGMA 2002-D19~\cite{agma2002}:
\begin{itemize}[nosep]
\item Aim for contact near the middle of the usable involute, midway between $d_{\rm Lf}$ and $d_{\rm La}$ (its 9.4.8). There the
  gain stays between \pinMidGain{} and \pinMidGainMax{} over the envelope. The standard $1.728m_n$ pin contacts the involute,
  clears the root and projects past the tip for every external gear we studied; the $1.68m_n$ internal pin does the same
  for every internal gear.
\item Avoid pins sized for contact at the TIF. In our envelope such a pin never projects past the tip, and in
  \pinLowRootPct\% of gears it would also touch the root unless relieved.
\item Allow for tooth thickness. At the low end the contact diameter moves radially by up to \pinDrcLow{} times any change in tooth thickness, so a tooth at the thin end of its tolerance can push a low contact into the fillet.
\item Avoid the very top as well. High contact has the lowest gain, but it sits at the edge of the tip break or chamfer,
  where the standard warns of false readings (9.1).
\item On internal gears the high-gain end is the minor (tip) side, where pins are easiest to fit (gain up to
  \intTipGainMax); the $1.44m_n$ internal pin never projects inside the minor circle with our proportions.
\item For helical gears with an odd tooth count, use balls or the standard's special two-pin methods (9.1, 9.8, 9.9).
\end{itemize}

\paragraph{How an error becomes a length.} An error in $\inv\phi$, or in the angle $\phi$ found from it, turns into a
length through simple derivatives. Since $dy/d\phi=\tan^2\phi$, an error $\delta y$ in the input of the inverse becomes
an angle error $\delta\phi=\delta y/\tan^2\phi$. The angle error is largest where $\phi$ is small, low on the flank. The inverse and
forward functions enter inspection in four places.
\begin{itemize}[nosep]
\item \emph{Profile inspection.} A CMM or gear checker measures the flank against roll angle or roll length, or as radius
  and polar angle. With polar data the deviation along the line of action is $f=r_b\,(\theta_{\rm meas}-\theta_{\rm nom}(R))$.
  The nominal angle $\theta_{\rm nom}$ contains $\inv\phi(R)$ from~\eqref{eq:halfangle}, so an error $\delta$ in $\inv\phi$
  enters $f$ as $r_b\,\delta$. The deviations are then reduced to $F_\alpha$, $f_{f\alpha}$ and $f_{H\alpha}$
  (ISO~1328-1~\cite{iso1328}, AGMA~2015-1~\cite{agma2015}) over a range that normally starts near the SAP, low on the
  flank.
\item \emph{Tooth thickness at a diameter} needs only the forward function: $\delta s_y=d_y\,\delta(\inv\phi)$.
\item \emph{Over pins or balls.} For an even tooth count and pin diameter $d_p$, the pressure angle $\phi_M$ at the pin
  centre is found with the inverse from
  \begin{equation}\label{eq:pins}
  \inv\phi_M=\frac{s}{mz}+\inv\alpha+\frac{d_p}{mz\cos\alpha}-\frac{\pi}{z},\qquad M=\frac{mz\cos\alpha}{\cos\phi_M}+d_p,
  \end{equation}
  where $s$ is the reference tooth thickness. An angle error moves $M$ at the rate
  $\partial M/\partial\phi_M=d_b\sin\phi_M/\cos^2\phi_M$. Span measurement, by contrast, uses only the forward function at $\alpha$.
\item \emph{Centre distance.} For a shifted pair the working pressure angle comes from the inverse of
  $\inv\alpha_w=\inv\alpha+2\tan\alpha\,(x_1+x_2)/(z_1+z_2)$, and then $a_w=a\cos\alpha/\cos\alpha_w$.
\end{itemize}

\paragraph{Tolerances used for scale.} We compare errors with the profile tolerances of ISO~1328-1:2013 (ANSI/AGMA
ISO~1328-1-B14, reaffirmed 2024)~\cite{iso1328}. For class $A$, with $m_n$ and $d$ in millimetres, they are
\begin{gather*}
f_{H\alpha T}=(0.4m_n+0.001d+4)\,\sqrt2^{\,A-5},\qquad f_{f\alpha T}=(0.55m_n+5)\,\sqrt2^{\,A-5},\\
F_{\alpha T}=\sqrt{f_{H\alpha T}^2+f_{f\alpha T}^2}\qquad(\text{all in }\mu\mathrm{m}),
\end{gather*}
rounded as the standard prescribes (Appendix~\ref{app:iso}). For $m=2$~mm and $d=40$~mm, class 5 gives
$f_{f\alpha T}=\isoFfFiveA$~$\mu$m and $F_{\alpha T}=\isoFaFiveA$~$\mu$m, and class 11 gives $F_{\alpha T}=\isoFaElevenA$~$\mu$m.
An error in $M$ or in centre distance is not a profile deviation, so the comparison only shows when an error becomes comparable to the gear's accuracy class.

\paragraph{Results.} Table~\ref{tab:pins} applies~\eqref{eq:pins} to gears from $d=20$~mm to $d=1$~m with $d_p=1.728m$,
for five ways of computing the inverse: this work; a naive Newton iteration in binary64 (Appendix~\ref{app:gears}; not the textbook Newton of Section~\ref{sec:invbench}); the same in binary32; Cheng's
explicit approximation~\cite{cheng1992} used without iteration; and the leading term $(3y)^{1/3}$.

\begin{table}[ht]
\centering\small
\input{tables/gear_pins.tex}
\caption{Error in the measurement over pins $M$ caused only by the inverse involute, $\alpha=20^\circ$, $x=0$,
$d_p=1.728m$, with ISO~1328-1:2013 form tolerances for scale. Generated by the author's script.}
\label{tab:pins}
\end{table}

\paragraph{What the numbers say.} Naive \emph{binary64} keeps
enough absolute accuracy that its length errors are below $10^{-10}$~$\mu$m in every case here, at most
\pinSixtyFourMax~$\mu$m over pins and \fwSixtyFourMax~$\mu$m in thickness (Table~\ref{tab:gearfwd}). Two shortcuts do cost micrometres:
\begin{itemize}[nosep]
\item \emph{Approximations used in place of a converged inverse.} Cheng's explicit formula costs \pinChengMin{} to
  \pinChengMax~$\mu$m over pins, more than the class-\chengExcMin{} form tolerance of the smallest gear and the
  class-\chengExcMax{} form tolerance of the 1~m gear. For the shifted pair $m=10$~mm, $z_1=z_2=50$, $x_1+x_2=0.5$
  ($a_w=\cdAw$~mm) it puts \cdCheng~$\mu$m into the centre distance, more than the class-\cdChengExc{} form tolerance; binary32
  and binary64 Newton give \cdThirtyTwo{} and \cdSixtyFour~$\mu$m. The leading term alone costs hundreds to thousands of
  micrometres.
\item \emph{Short formats.} Binary32 Newton costs at most about \pinThirtyTwoMax~$\mu$m over pins, under every class-1 tolerance
  here. Forward evaluation in binary16, as on some accelerators, costs \fwSixteenFive~$\mu$m in thickness at
  $\phi=5^\circ$ on a 1~m gear, more than its class-\fwSixteenExc{} form tolerance.
\end{itemize}
A faithfully rounded binary64 inverse therefore helps inspection in two ways: it removes the approximation error, and it gives the same last bit on every machine (Section~\ref{sec:repro}), so two inspection systems that evaluate the same nominal geometry agree exactly. Whether the lost relative digits of
naive binary64 matter elsewhere, for example in differences of nearby $\inv$ values or in chained solvers, depends on
the computation and is not quantified here.

\begin{figure}[tbp]
\centering
\includegraphics[width=0.66\textwidth]{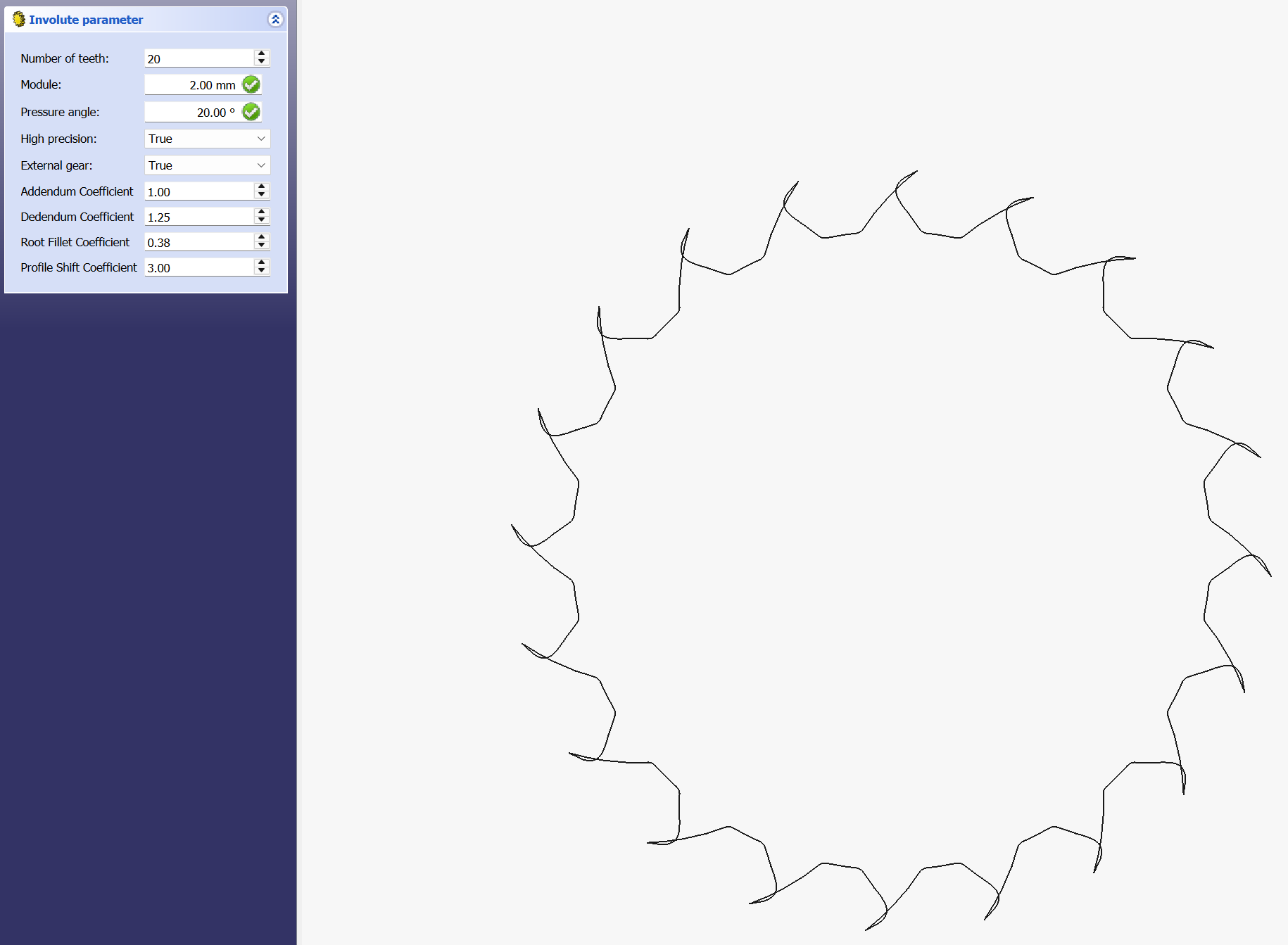}\\[-0.3em]
{\small (a) screen view}\\[0.8em]
\begin{minipage}[t]{0.42\textwidth}\centering
\includegraphics[width=\textwidth]{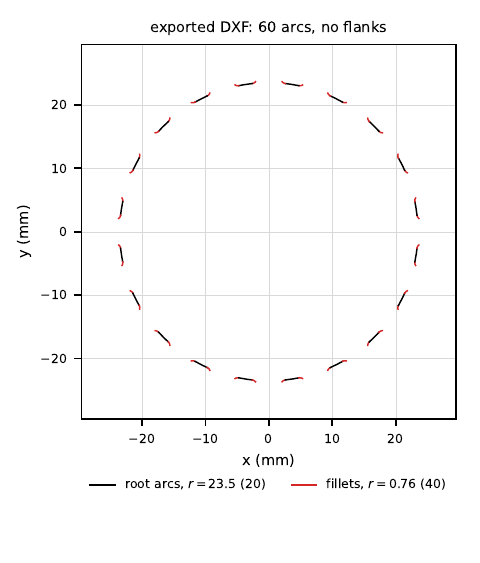}\\[-0.3em]
{\small (b) exported DXF}
\end{minipage}\hfill
\begin{minipage}[t]{0.42\textwidth}\centering
\includegraphics[width=\textwidth]{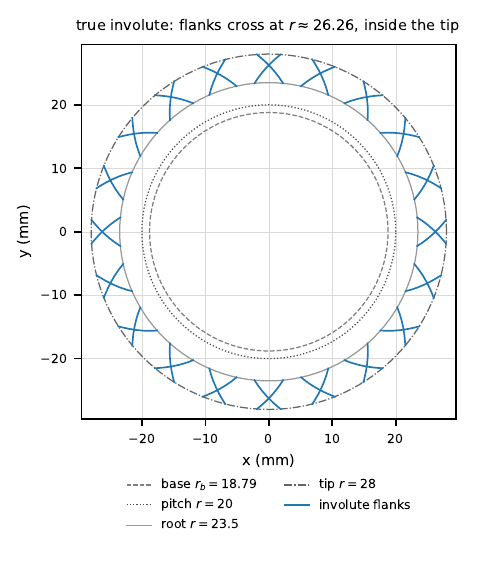}\\[-0.3em]
{\small (c) true geometry}
\end{minipage}
\caption{The worked example ($z=20$, $m=2$~mm, $\alpha=20^\circ$, addendum 1.00, dedendum 1.25, root fillet 0.38) with an
out-of-range profile shift of 3.00 in FreeCAD~1.0.1 (PartDesign Involute gear). (a)~The screen view: the value is accepted
without warning and a shape is drawn, but the flanks cross and the teeth self-intersect. (b)~The DXF exported for the same
gear: 60 arcs only (20 root arcs, $r=23.5$~mm, and 40 fillets, $r=0.76$~mm); every flank is missing. (c)~The true geometry
from the forward involute: base ($r_b=18.79$~mm), pitch (20~mm), root (23.5~mm) and tip (28~mm) circles, and involute
flanks that cross at $r\approx26.26$~mm, inside the tip. By standard geometry this tooth becomes pointed at $x\approx1.22$.
Panels (b) and (c) are drawn at the same scale. Shown as an example of missing input checks.}
\label{fig:freecad}
\end{figure}

\paragraph{Missing input checks.}\label{sec:inputchecks} Figure~\ref{fig:freecad} shows the worked example of this paper ($z=20$, $m=2$~mm, $\alpha=20^\circ$,
addendum $1.00m$, dedendum $1.25m$, root fillet $0.38m$) entered in a common open-source CAD tool with a profile shift of
$x=3.00$. The tool accepts the value and still draws a shape, but the flanks cross and the teeth become self-intersecting
spikes (Figure~\ref{fig:freecad}a). The DXF file the tool exports for the same gear holds only root arcs and fillets,
20 arcs of radius 23.5~mm and 40 of radius 0.76~mm, with every flank dropped (Figure~\ref{fig:freecad}b). The screen and the file disagree, neither is a valid gear, and a shop or CAM program that receives the file gets no warning. From standard textbook geometry (forward involute only, tip diameter $d+2(1+x)m$, Eq.~\eqref{eq:halfangle} at the
tip), this gear's tooth is already pointed at $x\approx1.22$ (tip diameter about 48.9~mm); at $x=0$ its tip thickness is
\thTip~mm. A shift of 3.00 is far past that limit: the true flanks cross at a radius of about 26.26~mm, inside the
28~mm tip circle (Figure~\ref{fig:freecad}c).

Gear software should warn, or refuse, when the inputs give a pointed tooth, an undercut,
crossed flanks or an addendum that does not fit, and it should never silently draw or export a broken shape. The basic
checks are simple:
\begin{itemize}[nosep]
\item tip thickness greater than zero;
\item no undercut, and no crossing of the two flanks of a tooth;
\item tip and root diameters within their limits (the addendum fits, and the root does not cut into the involute that
  is needed).
\end{itemize}

\subsection{Over pins and balls: where the contact lies}\label{sec:pinsloc}
ANSI/AGMA 2002-D19~\cite{agma2002} lets the designer choose the pin or ball diameter $D_d$, provided the contact lies on
the unmodified involute between the root and tip involute limit diameters $d_{\rm Lf}$ and $d_{\rm La}$ (its 9.4.1, with
the limits $D_{d\,\rm max}$ and $D_{d\,\rm Lf}$ of 9.4.3 and 9.4.4). Depending on the instrument, the pin must also
project past the tip (9.4.5), and it must clear the root unless it is relieved (9.4.6). Clause 9.4.8 recommends contact
midway between $d_{\rm Lf}$ and $d_{\rm La}$ and lists the commercial pin series, $1.728/P$ for most external gears
($1.728m_n$ in module terms). We followed those rules over the envelope (\pinGears{} external spur gears without
undercut; \pinExcl{} undercut gears excluded) and for helical gears measured with balls. For each gear we took five
contact positions: at $d_{\rm Lf}$ (the TIF), the lowest contact whose pin projects past the tip, the 9.4.8 midpoint, the
$1.728m_n$ pin, and at $d_{\rm La}$. The pin-centre diameter follows from the inverse involute of $y=\inv\alpha_k$ (Eq.~\eqref{eq:pins}, written as in clause 9.5 of ANSI/AGMA 2002-D19); $M$ follows from its clause 9.7.2, with a factor $\cos(\pi/2z)$ for odd $z$.
Appendix~\ref{app:pins} lists the assumptions, including a $0.05m$ tip break that sets $d_{\rm La}$.

Two derivatives govern the study. For a fixed pin, $\partial M/\partial s_n=k\cos\alpha_t/(\cos\beta\sin\alpha_k)$ with
$k=1$ for even $z$ and $k=\cos(\pi/2z)$ for odd $z$. A normal form deviation $e$ on the contacted flanks therefore moves
$M$ by $\textrm{gain}\cdot e$, with $\textrm{gain}=2k\cos\alpha_t/(\cos\beta\cos\alpha_n\sin\alpha_k)$, which is $2/\sin\alpha_k$ for an
even spur gear. An error $\delta\alpha_k$ in the inverse moves $M$ by $\partial M/\partial\alpha_k=2kr_b\sin\alpha_k/\cos^2\alpha_k$.
Table~\ref{tab:pinsloc} gives the results for four gears, and Figure~\ref{fig:pins} in the appendix plots them along the
flank.

\begin{table}[ht]
\centering\scriptsize\setlength{\tabcolsep}{3pt}
\input{tables/pins_contact.tex}
\caption{Measurement over two pins or balls against contact position (ANSI/AGMA 2002-D19 rules; nominal tooth
thickness). $\alpha_k$: transverse pressure angle at the pin centre; $y=\inv\alpha_k$, the inverse-involute input. Gain: change
in $M$ per unit normal form deviation on the contacted flanks; the next column applies it to a local deviation equal to
the ISO~1328-1:2013 class-5 $f_{f\alpha T}$. Cheng, b32, this work: error in $M$ from the inverse alone (``this work'' is
the larger of the solver's error and the error from forming $y$ in binary64). Proj.: the pin projects past the tip;
root: it clears the root. Generated by the author's script.}
\label{tab:pinsloc}
\end{table}

\paragraph{Which contact is risky, and why.} The numbers confirm the common view that low contact is riskier, for physical rather than numerical reasons.

\emph{Numerical sources.} The inverse is evaluated at the pin-centre angle $\alpha_k$, not at the local pressure angle of
  the contact point, and $\alpha_k$ is always larger than that local angle. With contact at the TIF, $\alpha_k$ is still at least
  \pinLowAkMin$^\circ$ over the envelope ($y\ge\pinLowYMin$), and any pin that projects past the tip has $\alpha_k\ge\pinProjAkMin^\circ$. This
  work, naive binary64 Newton, and forming $y$ in binary64 all stay below \pinConvMax~$\mu$m per millimetre of module at every
  position. Binary32 Newton stays below \pinBtwoMax~$\mu$m per millimetre of module, under every class-1 tolerance up to
  $m=10$. The approximations go the other way. Cheng's formula and the leading term are most accurate as $y\to0$. So their errors tend to be \emph{smallest} at the low contact and largest at the high one: \exLowCheng{} to \exHighCheng~$\mu$m from TIF to $d_{\rm La}$
  on the $z=20$ gear, and up to \pinChengHighMax~$\mu$m per millimetre of module at $d_{\rm La}$ over the envelope ($z=\pinChengHighZ$,
  $\alpha=\pinChengHighA^\circ$). Cheng's error also grows with $z$, because $\partial M/\partial\alpha_k$ scales with $r_b$.

\emph{Physical sources.} The gain at a low contact (\exLowGain{} on the $z=20$ gear against \exMidGain{} at the
  midpoint, \pinLowGainMax{} over the envelope) and the pin-fit limits are summarised at the start of this section. The
  contact also moves radially with tooth thickness, by up to \pinDrcLow{} times the thickness change at the low end and
  \pinDrcHigh{} at the high end. At the high end the gain is lowest, but the contact sits by construction at the edge of
  the tip break or chamfer, where the same thickness effect can push it onto the modified surface (a false reading, 9.1).
  Odd tooth counts change the geometry only mildly: the factor $\cos(\pi/2z)$ is at least \pinKoddMin{} (at $z=5$).

\noindent The midpoint and the $1.728m_n$ pin are the well-conditioned choices. With a converged inverse, the arithmetic
contributes nothing measurable anywhere on the flank. With an explicit approximation, the arithmetic error alone can exceed
the class-5 form tolerance, and on the $m=10$ gear even the class-11 one; it is worst high on the flank.

\paragraph{Internal gears.} The standard treats internal gears with $z<0$, and they are measured between pins or balls.
We repeated the study for internal spur gears, plus one internal helical gear measured with balls (Appendix~\ref{app:pinsint}). The
model uses a pinion-type cutter, the root form diameter of ISO~21771-1:2024~\cite{iso21771}, and its interference checks
against the cutter and the mating pinion. The envelope then holds \intGears{} internal gears, from a smallest $|z|$ of \intNminLo{} (at $28^\circ$) to \intNminHi{} (at $30^\circ$), depending on $\alpha$, up to $|z|=500$. On an internal gear the pin centre sits \emph{nearer} the base circle than the contact, so the inverse
is evaluated below the local pressure angle. The difference is up to \intTipLagMax$^\circ$ at the minor (tip) side, where
the low local pressure angles now lie. Even so, $y$ stays at or above \intTipYMin, inside the range the external gears already
reach. The arithmetic picture is unchanged: a converged inverse contributes less than \intConvMax~$\mu$m per millimetre of
module, and Cheng's formula is again worst at the large-pressure-angle end, which on an internal gear is the root (up to \intRootCheng~$\mu$m per millimetre of module). The physics does change. The form gain peaks at the minor side (up to \intTipGainMax), and so does pin projection: the
large minor-side pins always project, while no root-limit pin does. The $1.68m_n$ internal pin of 9.4.8 is valid for every
gear, but the $1.44m_n$ pin never projects inside the minor circle with these proportions.

\section{Manufacturing and CAD}\label{sec:mfg}
How the involute reaches the part depends on the process; Appendix~\ref{app:mfg} gives the detail.
\emph{Hobs and shaper cutters} produce the involute by generating motion, so tool errors, not arithmetic, set the
profile. A hob pressure-angle error of one arc-minute tilts the profile of the worked example by \hobFhTwo~$\mu$m, under the
class-1 $f_{H\alpha T}$ of \hobFhOneTwoIso~$\mu$m; five arc-minutes exceed class \hobExcFiveTwo, and for $m=10$~mm, $z=100$ one
arc-minute (\hobFhTen~$\mu$m) already exceeds class \hobExcOneTen. \emph{Wire EDM} from a \emph{DXF} or \emph{STEP file} copies the drawing one to one, so every flank point must come from the true involute through~\eqref{eq:halfangle}; the number of
points needed is small (\dxPolyOne{} polyline points hold the worked example's flank to 1~$\mu$m). \emph{3D printing}
adds binary32 STL quantisation (\stlSpacingSmall~$\mu$m at 22~mm, negligible) and coarse process scatter, but a wrong
source profile is a systematic error in every tooth.

\paragraph{Circular-arc teeth.} The SolidWorks documentation states that SolidWorks Toolbox gears ``are not true involute gears that you can
use for manufacturing''~\cite{swtoolbox}, and published descriptions give the flank as a single circular arc of radius
$d/5$, centred on the base circle and through the pitch point~\cite{hawkridge2024,ju_gears}. We measured a SolidWorks~2026 Toolbox export (ISO spur gear, $m=2$~mm, $z=20$, $\alpha=20^\circ$, face width 12~mm, STEP AP214~\cite{toolbox2026}) by reading the exact surface definitions in its B-rep and cross-checking them with OpenCASCADE. Each flank is one cylindrical face of radius \tbR~mm
($=d/5$) whose axis lies on the base circle (within \tbRcErr~mm). The tip is a cylinder of diameter \tbDa~mm and the root
a cylinder of diameter \tbDf~mm. The flank arc meets the root circle at a sharp corner, with no root fillet. Both
flanks and all 20 teeth are identical. The one difference from the published model is that the arc centres are rotated
\tbRot$^\circ$ further from the tooth centreline ($\tbOff^\circ$ instead of $\tbOffModel^\circ$), so the arc does not pass
through the pitch point. The tooth is \tbSpitch~mm thick at the pitch circle instead of $\pi m/2$, which builds in
\tbThin~mm of circumferential backlash per tooth (\tbNormOff~mm normal per flank).

The SolidWorks Toolbox sketch shows that the thinning is intentional. Its tooth-half dimension is
$\mathrm{Half\_CT}=\tfrac{d}{2}\sin(90^\circ/z)-\tfrac{7}{4}\,\mathrm{Backlash}$, where $d=zm$ is the pitch diameter. For $z=20$ this gives
$d\sin(4.5^\circ)/2=\tbHalfNom$~mm. The sketch shows \tbHalfCT~mm, so the backlash entry is \tbBL~mm. The full chordal
thickness is then $2(\tbHalfNom)-3.5\times\tbBL=\tbChordFull$~mm, which agrees with the measured \tbSpitch~mm (the
arc is longer than its chord by about 0.003~mm at this size). So each tooth is thinned by 3.5 times the backlash entry.
The same thinning moves the arc off the pitch point. The model therefore has backlash built into every tooth; allow for it when you use the model as a CAD reference or for inspection.

Between the SAP of a mating $z=20$ gear and the tip, the measured flank lies \tbDevMin{} to \tbDevMax~$\mu$m inside the
involute: a span of \tbSpan~$\mu$m (from the unrounded end values), against \arcRange~$\mu$m for the published model. Evaluated as ISO~1328-1:2013
prescribes, with a least-squares mean line over $L_\alpha=\tbLa$~mm (Appendix~\ref{app:iso}), the uniform thinning drops
out. The total profile deviation $F_\alpha=\tbFa$~$\mu$m meets class~\tbFaCls{} ($F_{\alpha T}=\tbFaTol$~$\mu$m) and the form
deviation $f_{f\alpha}=\tbFf$~$\mu$m meets class~\tbFfCls{} (\tbFfTol~$\mu$m). The slope deviation $|f_{H\alpha}|=\tbFh$~$\mu$m
lies just outside class~11 (\tbFhElevenTol~$\mu$m). The published model gives $\tbModelFh$~$\mu$m, just inside class~\tbModelFhCls,
so the slope class depends on sub-micrometre detail. The best possible single arc over the SAP-to-tip range spans \arcBand~$\mu$m (within $\pm\arcBest$~$\mu$m, Appendix~\ref{app:mfg}, where the half-width is rounded on its own), which meets class~\arcBestFitA{} ($F_{\alpha T}=\isoFaBestA$~$\mu$m) on that span. For $z=12$, using
the published model (we did not export a $z=12$ part), the span is \arcRangeB~$\mu$m and the best arc spans \arcBandB~$\mu$m
(class~\arcBestFitB). In a $z=20$ pair of measured flanks, the instantaneous ratio swings between \tbRatioMin{} and
\tbRatioMax{} instead of staying at 1. The swing depends only on the base radius, the arc radius and the centre distance,
so it is the same as for the published model. Arc teeth therefore give misleading results in meshing, transmission-error and contact studies, and printed or EDM-cut arc gears will not run at a constant ratio.

%% file: tables/gear_pins.tex
\begin{tabular}{lrrrr}
\toprule
gear & $m=1$, $z=20$ & $m=2$, $z=20$ & $m=2$, $z=40$ & $m=10$, $z=100$ \\
$d$ / pin $d_p$ (mm) & 20 / 1.728 & 40 / 3.456 & 80 / 3.456 & 1000 / 17.28 \\
$\phi_M$ ($^\circ$) & 24.552 & 24.552 & 22.531 & 21.093 \\
$M$ (mm) & 22.3900 & 44.7800 & 84.8434 & 1024.4564 \\
$\partial M/\partial\phi_M$ (mm/rad) & 9.44 & 18.88 & 33.76 & 388.50 \\
\midrule
\multicolumn{5}{l}{\emph{error in $M$ ($\mu$m) from the inverse involute alone}} \\
this work & $2.5\times10^{-13}$ & $5.0\times10^{-13}$ & $5.8\times10^{-13}$ & $6.0\times10^{-12}$ \\
naive Newton, binary64 & $7.8\times10^{-13}$ & $1.6\times10^{-12}$ & $5.0\times10^{-12}$ & $4.9\times10^{-11}$ \\
naive Newton, binary32 & $4.6\times10^{-4}$ & $9.1\times10^{-4}$ & $6.2\times10^{-4}$ & 0.023 \\
Cheng explicit & 2.5 & 4.9 & 5.7 & 42.5 \\
$(3y)^{1/3}$ only & 104 & 209 & 286 & 2{,}684 \\
\midrule
\multicolumn{5}{l}{\emph{ISO 1328-1:2013 $f_{f\alpha T}$ ($\mu$m, rounded per the standard), for scale}} \\
class 2 & 2 & 2.2 & 2.2 & 3.7 \\
class 5 & 5.5 & 6 & 6 & 11 \\
class 8 & 16 & 17 & 17 & 30 \\
class 11 & 44 & 49 & 49 & 84 \\
\bottomrule
\end{tabular}

%% file: tables/pins_contact.tex
\begin{tabular}{lrrrrrrrrrcc}
\toprule
contact & $d_c$ & $D_d$ & $\alpha_k$ & $y$ & gain & gain$\cdot f_{f\alpha T}^{(5)}$ & Cheng & b32 & this work & proj. & root \\
 & (mm) & (mm) & ($^\circ$) & & & ($\mu$m) & ($\mu$m) & ($\mu$m) & ($\mu$m) & & \\
\midrule
\multicolumn{12}{l}{\emph{spur, $m=2$, $z=20$, $\alpha=20^\circ$ (pins)}; $f_{f\alpha T}$ class 2/5/8: 2.2/6/17~$\mu$m} \\
$d_{\rm Lf}$ (TIF) & 37.64 & 2.412 & 6.7 & 5.30e-4 & 17.2 & 103 & 0.42 & $1.3\times10^{-3}$ & $9.5\times10^{-13}$ & -- & \checkmark \\
lowest proj. & 39.63 & 3.232 & 22.8 & 2.24e-2 & 5.2 & 31 & 3.1 & $6.2\times10^{-4}$ & $1.5\times10^{-12}$ & \checkmark & \checkmark \\
midpoint & 40.72 & 3.921 & 27.5 & 4.07e-2 & 4.3 & 26 & 8.8 & $6.2\times10^{-4}$ & $3.6\times10^{-13}$ & \checkmark & \checkmark \\
$1.728m_n$ & 40.01 & 3.456 & 24.6 & 2.83e-2 & 4.8 & 29 & 4.9 & $9.1\times10^{-4}$ & $1.6\times10^{-12}$ & \checkmark & \checkmark \\
$d_{\rm La}$ & 43.80 & 6.798 & 37.9 & 1.17e-1 & 3.3 & 20 & 15 & $8.0\times10^{-4}$ & $1.0\times10^{-12}$ & \checkmark & \checkmark \\
\addlinespace
\multicolumn{12}{l}{\emph{spur, $m=2$, $z=40$, $\alpha=14.5^\circ$ (pins)}; $f_{f\alpha T}$ class 2/5/8: 2.2/6/17~$\mu$m} \\
$d_{\rm Lf}$ (TIF) & 77.59 & 2.633 & 5.3 & 2.71e-4 & 21.5 & 129 & 0.59 & $2.8\times10^{-3}$ & $1.9\times10^{-11}$ & -- & \checkmark \\
lowest proj. & 79.90 & 3.244 & 16.4 & 8.15e-3 & 7.1 & 42 & 0.91 & $3.5\times10^{-3}$ & $5.7\times10^{-12}$ & \checkmark & \checkmark \\
midpoint & 80.69 & 3.547 & 18.7 & 1.21e-2 & 6.2 & 37 & 0.65 & $7.7\times10^{-4}$ & $4.9\times10^{-12}$ & \checkmark & \checkmark \\
$1.728m_n$ & 80.47 & 3.456 & 18.1 & 1.09e-2 & 6.4 & 39 & 0.12 & $1.2\times10^{-4}$ & $5.8\times10^{-12}$ & \checkmark & \checkmark \\
$d_{\rm La}$ & 83.80 & 5.115 & 25.6 & 3.23e-2 & 4.6 & 28 & 13 & $4.0\times10^{-3}$ & $3.8\times10^{-12}$ & \checkmark & \checkmark \\
\addlinespace
\multicolumn{12}{l}{\emph{spur, $m=10$, $z=100$, $\alpha=20^\circ$ (pins)}; $f_{f\alpha T}$ class 2/5/8: 3.7/11/30~$\mu$m} \\
$d_{\rm Lf}$ (TIF) & 981.54 & 9.759 & 17.3 & 9.58e-3 & 6.7 & 74 & 4.9 & 0.023 & $6.1\times10^{-12}$ & -- & -- \\
lowest proj. & 998.34 & 16.110 & 20.6 & 1.63e-2 & 5.7 & 63 & 34 & 0.030 & $1.0\times10^{-11}$ & \checkmark & \checkmark \\
midpoint & 1000.27 & 16.929 & 20.9 & 1.72e-2 & 5.6 & 62 & 40 & 0.054 & $5.8\times10^{-12}$ & \checkmark & \checkmark \\
$1.728m_n$ & 1001.09 & 17.280 & 21.1 & 1.76e-2 & 5.6 & 61 & 43 & 0.023 & $1.1\times10^{-11}$ & \checkmark & \checkmark \\
$d_{\rm La}$ & 1019.00 & 25.768 & 24.1 & 2.66e-2 & 4.9 & 54 & 109 & 0.031 & $2.2\times10^{-11}$ & \checkmark & \checkmark \\
\addlinespace
\multicolumn{12}{l}{\emph{helical, $m_n=2$, $z=21$ (odd), $\alpha_n=20^\circ$, $\beta=30^\circ$ (balls)}; $f_{f\alpha T}$ class 2/5/8: 2.2/6/17~$\mu$m} \\
$d_{\rm Lf}$ (TIF) & 45.50 & 2.230 & 13.1 & 4.10e-3 & 9.9 & 60 & 1.0 & $8.9\times10^{-4}$ & $8.1\times10^{-13}$ & -- & \checkmark \\
lowest proj. & 48.14 & 3.228 & 24.8 & 2.94e-2 & 5.4 & 32 & 6.2 & $1.6\times10^{-4}$ & $2.6\times10^{-13}$ & \checkmark & \checkmark \\
midpoint & 48.90 & 3.619 & 27.2 & 3.93e-2 & 4.9 & 30 & 9.9 & $3.5\times10^{-3}$ & $3.4\times10^{-13}$ & \checkmark & \checkmark \\
$1.728m_n$ & 48.59 & 3.456 & 26.3 & 3.52e-2 & 5.1 & 31 & 8.4 & $1.3\times10^{-5}$ & $1.4\times10^{-12}$ & \checkmark & \checkmark \\
$d_{\rm La}$ & 52.30 & 5.901 & 35.9 & 9.71e-2 & 3.9 & 23 & 21 & $3.2\times10^{-3}$ & $1.5\times10^{-12}$ & \checkmark & \checkmark \\
\addlinespace
\bottomrule
\end{tabular}

%% file: app_history.tex
\section{A short history of the involute tooth}\label{app:history}
This appendix follows the history chapter of the author's video on the pressure angle~\cite{prestridge2023}. We checked each statement against the cited sources. Where the sources disagree with each other, or with the video, we
say so. Quotations from Euler were read from the page scans of the original printings held by the Euler Archive; the
section and page numbers are those of the originals.

\paragraph{Practice before theory.} Toothed wheels are far older than any theory of their teeth. Radzevich's survey
starts its pre-Eulerian period with the Antikythera mechanism, and names Leonardo da Vinci, D\"urer and Hooke among
those who took an interest in gears~\cite{radzevich2019}. Desargues, de La Hire and Camus came closest to a theory. In
1694 de La Hire published a treatise on epicycloids and their use in mechanics~\cite{delahire1694}. Litvin credits de La
Hire, Poncelet and Camus with developing cycloidal gearing~\cite{litvin1998}. Coy writes that de La Hire ``is said to
have'' also recommended the involute for gear teeth, about 1694~\cite{coy1983}; we have not found this in de La Hire's
own text, so we give it only as a report.

\paragraph{Camus.} Charles-\'Etienne-Louis Camus (1699--1768) read a memoir on the shape of the teeth of clock wheels
and pinions to the Paris Academy for the year 1733~\cite{camus1733}. The gear books of his \emph{Cours de math\'ematique}
(third part, 1751--1752) were translated into English in 1806~\cite{camus1806}. What is now called Camus's theorem
concerns teeth traced by a point of an auxiliary curve rolling on the two pitch curves, the basis of cycloidal
gearing~\cite{volkov2014,litvin1998}. Radzevich credits Camus with the first study of the law of conjugate action, but
argues that Camus did not state it correctly, because he allowed a curved line of action~\cite{radzevich2019}. Camus helped found the theory of tooth action, but the involute tooth is not his.

\paragraph{Euler, 1760: the involute set aside.} Leonhard Euler wrote \emph{De aptissima figura rotarum dentibus
tribuenda} (``on the most suitable shape for gear teeth''). The Euler Archive dates the writing to 1751; it was published
in 1760, in volume~5 of the St~Petersburg \emph{Novi Commentarii}, for 1754--1755~\cite{euler1760}. In \S19 (p.~314) one
case of his general solution gives a tooth curve ``ex euolutione circuli'', from the unwinding of a circle. In \S20
(p.~314) he rejects it: the two curves would not only touch at the contact point but also cross there, so that the teeth
``se mutuo penetrare deberent'' (would have to penetrate each other). In \S\S21--22 (pp.~314--315) he allows such teeth
only when a very large wheel works with a small one, and advises against them for wheels of nearly equal size. The
circles in \S20 are drawn about the wheel centres through the point~$C$ on the line of centres, which we read as the pitch
circles. If that reading is right, this is a zero-obliquity case and not the base-circle involute in use today.

\paragraph{Euler, 1767: the involute proposed.} Euler's second paper on the subject is headed \emph{Supplementum. De figura
dentium rotarum}; it was read in Berlin in 1762 and published in 1767 in volume~11 of the \emph{Novi Commentarii},
pp.~207--231~\cite{euler1767}. A line $RS$ through the contact region meets the line of centres $AB$ at an angle~$\omega$,
and $AR$, $BS$ are the perpendiculars to it from the wheel centres. Euler shows that $\omega$ must not be a right angle
but moderately acute or obtuse (\S20, p.~221), and that the teeth have enough thickness only if $\omega>60^\circ$
(\S\S25--26, pp.~225--226); his worked construction draws $RS$ at $60^\circ$ ``or somewhat more'' to $AB$ (\S28,
p.~228). In modern terms the pressure angle is $90^\circ-\omega$, so this asks for pressure angles below $30^\circ$. He
first makes each flank a circular arc centred at $R$ or $S$, the case he calls ``simplicissimum'' (\S24, p.~224), and then
notes that it is only approximately right. Under the heading ``Accuratior determinatio figurae dentium'' (\S29,
pp.~228--229) he finds that if $\omega$ is held constant, the perpendiculars $AR$ and $BS$ are constant too, ``ideoque
utraque curua EOM et FON ex euolutione circuli nata'': both tooth curves are involutes, of the circles of radius $AR$
about~$A$ and $BS$ about~$B$, the base circles of today's terminology. Because these curves ``cum sint descriptae
facillime, solutionem commodissimam praebere videntur'' (being very easily drawn, seem to give the most convenient
solution). In \S31 (p.~230) he gives a construction that wraps a line onto the inner circle and unwinds it. This is the
involute system as it is used now, with a constant pressure angle. Radzevich likewise finds the involute tooth, and a
formula equivalent to the Euler--Savary equation, in this second paper~\cite{radzevich2019}. Volkov and Babichev
credit Euler with finding the special properties of the circle involute and proposing it for gearing~\cite{volkov2014}.
Euler's circular-arc approximation of \S24 is the construction that Willis later turned into a rule (below)
and a close relative of the circular-arc flanks that Section~\ref{sec:mfg} measures in CAD tooth models.

\paragraph{Savary.} F\'elix Savary (1797--1841) was a French astronomer who studied and later taught at the \'Ecole
polytechnique, where he became professor of astronomy and geodesy in 1831~\cite{mactutorsavary}. Continuing Euler's work,
he found the relation between the radii of curvature of conjugate profiles and those of the pitch curves (the centrodes),
now known as the Euler--Savary equation~\cite{volkov2014}. Bobillier later based a graphical construction for the
centres of curvature on it~\cite{volkov2014}. Radzevich holds that the fundamental theorem of parallel-axis gearing was
known in its final form to Euler and Savary. He proposes calling it the Camus--Euler--Savary theorem, or ``CES
theorem'' for short, in place of the usual ``Willis theorem''~\cite{radzevich2019}. This is the ``CES'' of the video.

\paragraph{Willis.} Robert Willis (1800--1875) of Cambridge published \emph{Principles of Mechanism} in
1841~\cite{willis1841}, and the fundamental theorem became widely known through it~\cite{radzevich2019}. For teeth whose
flank is a single circular arc, Willis struck the arc through the pitch point from a centre on a line through the pitch
point, at the foot of the perpendicular from the wheel centre: Euler's \S24 construction. He showed that the centre's
distance from the pitch point is $R\cos\theta$, where $R$ is the pitch radius and $\theta$ is the angle between that line
and the line of centres. Holding $\theta$ fixed lets any two wheels of a set work together. He chose $\theta=75^\circ30'$
``which is a very convenient value'', because $\cos75^\circ30'=0.25038\approx\tfrac14$~\cite[Arts.~137--138]{willis1841}.
In modern terms this is a pressure angle of $14.5^\circ$. The arc centre lies on the base circle, so the arc is the circle
of curvature of the involute at the pitch point. For his more accurate two-arc teeth and the odontograph that draws them,
Willis used $75^\circ$ instead, ``by various trials'' the angle that gave the best form~\cite[Arts.~139--142]{willis1841}.
Willis's $75^\circ$ matches Wilfred Lewis's recollection that the involute systems in use when he took up the subject were the
``standard'' $14\tfrac12^\circ$ system and a separate $15^\circ$ system recommended by Willis (remarks of 1900, known to us only through Crosher~\cite{crosher2013}; see below).
None of our sources says that the $14.5^\circ$ standard was taken from Willis.

\paragraph{Brown \& Sharpe and the $14.5^\circ$ angle.} Brown \& Sharpe's 1886 \emph{Practical Treatise on Gearing} uses
the same quarter-radius construction. A point one-fourth of the pitch radius from the pitch point gives ``lines of
pressure'' at $75\tfrac12^\circ$ to the radius, and the treatise notes that a rack with the usual addendum starts to
interfere with gears of 31 teeth~\cite[Chs.~IV, XI]{brownsharpe1886}. Lewis recalled recommending a $20^\circ$ system
instead, which William Sellers \& Co.\ adopted~\cite{crosher2013}.

\paragraph{From $14.5^\circ$ to $20^\circ$.} The American standard ASA B6.1-1932 covered a $14\tfrac12^\circ$
full-depth form, a $20^\circ$ stub form and a $14\tfrac12^\circ$ composite form. Its successor, ANSI B6.1-1968, lists all
three as not recommended for new designs~\cite{machinery31}. In Germany, DIN~867 first appeared in July 1927 as
``Zahnform f\"ur Stirnr\"ader und Kegelr\"ader''~\cite{din867}. The September 1974 edition of DIN~867, which replaces the
September 1963 issue, specifies a basic-rack pressure angle $\alpha_P=20^\circ$ (clause~4 and its figure)~\cite{din867later}. The basic rack of
ISO~53:1998 has a $20^\circ$ pressure angle~\cite{iso53}. The video names $14.5^\circ$, $20^\circ$, $25^\circ$ and
$30^\circ$ as the four common angles. It notes that $20^\circ$ is the currently preferred angle, and that larger angles
give stronger but noisier teeth~\cite{prestridge2023}.

\paragraph{Points we could not confirm.}
\begin{itemize}[nosep]
\item \emph{Savary's own publication} of the curvature relation has not been found. The MacTutor biography names
  none~\cite{mactutorsavary}; it lists Dupont's study of the history of the Euler--Savary formula~\cite{dupont1963}, which
  we have not seen and give only as a pointer. The attribution here rests on Volkov and Babichev~\cite{volkov2014}.
\item \emph{De La Hire's recommendation of the involute} is known only at second hand~\cite{coy1983}.
\item \emph{Euler's two papers.} E330 opens by referring to the author's earlier work on tooth shape ``in volumine quarto''
  of the \emph{Novi Commentarii}, while E249 appeared in volume~5. We have not resolved this, so we do not state that E330
  is a supplement to E249. Our reading of the circles in E249 \S\S19--21 as pitch circles is an interpretation of the
  figures. The Euler Archive scan of E330 lacks p.~215, which none of the passages above depends on.
\item \emph{Willis, Brown and $14.5^\circ$.} The video says Willis chose $14.5^\circ$ in concert with Brown, of Brown \&
  Sharpe. We found no source for this, and Lewis describes the $14\tfrac12^\circ$ standard and Willis's $15^\circ$ as
  separate systems~\cite{crosher2013}. The video's description also says the angle may go back to about 1820;
  we found no source for that date and do not use it.
\item \emph{Lewis's remarks.} We quote Lewis only through the Gear Solutions column, which dates the remarks to an
  ASME meeting in 1900 and gives no page reference~\cite{crosher2013}. We have not consulted the primary source in
  the ASME \emph{Transactions}.
\item \emph{AGMA and $20^\circ$.} The video's description gives 1981 as the year AGMA made $20^\circ$ the preferred angle,
  and a secondary timeline gives 1980. We found no AGMA document fixing either date, so the appendix states neither.
\item \emph{DIN~867.} The 1927 edition is known to us only from the DIN Media catalogue~\cite{din867}. We have not seen
  its text, so we do not say what pressure angle it specified. The earliest edition we have seen is September
  1974~\cite{din867later}.
\item \emph{Video description.} It cites Euler's first paper as published in 1750 on pages 119--135. Those are the page
  numbers of the \emph{Opera Omnia} reprint, and the original appeared in 1760 on pages 299--316~\cite{euler1760}.
\end{itemize}

%% file: app_flank.tex
\section{Flank geometry of a spur gear tooth}\label{app:flank}
This appendix condenses the author's own notes on spur-gear flank geometry.

\paragraph{The involute.} The involute is the path of the end of a taut string unwrapping from the base circle. After
a roll angle $t$ the free length is the roll length $L=r_bt$, and the point lies at $R=r_b\sqrt{1+t^2}=\sqrt{r_b^2+L^2}$.
At that radius $\cos\phi=r_b/R$, $t=\tan\phi$ and $L=R\sin\phi$. The polar angle of the point, measured from where the
curve leaves the base circle, is $\inv\phi=\tan\phi-\phi$, which gives~\eqref{eq:halfangle}. Three steps locate any flank point: $\cos\phi=r_b/R$ gives $\phi$, then $\inv\phi$, then $\psi$ from~\eqref{eq:halfangle}.

\paragraph{Generation and the named points.} The tooth is generated, not drawn~\cite{litvin2004}. A cutter with a
straight-sided basic rack rolls against the blank, and the straight part of the rack flank produces the involute. The
rack's rounded tip produces a trochoidal fillet below it. The points that matter for rating and inspection are fixed by
roll lengths along the line of action, the common tangent of the two base circles. Going outward from the root:
\begin{itemize}[nosep]
\item the \emph{root}, cut by the rack tip, not by the involute;
\item the \emph{base circle}, where the involute starts geometrically. In the worked example it lies inside the fillet and
  is not part of the finished tooth;
\item the \emph{true involute form} (TIF, form diameter), where the cutter's straight flank stops cutting. The straight
  part reaches $h'=h_{a0}-\rho_{a0}(1-\sin\alpha)$ below the pitch line, so $L_{\rm TIF}=r\sin\alpha-h'/\sin\alpha$. A
  negative $L_{\rm TIF}$ means the cutter undercuts the tooth;
\item the \emph{start of active profile} (SAP), where the mate's tip first touches: $L_{\rm SAP}=a\sin\alpha-\sqrt{r_{a2}^2-r_{b2}^2}$.
  It must lie above the TIF, or the mate's tip runs into the fillet;
\item the \emph{lowest} and \emph{highest points of single tooth contact}, $L_{\rm LPSTC}=L_{\rm EAP}-p_b$ and
  $L_{\rm HPSTC}=L_{\rm SAP}+p_b$, with $L_{\rm EAP}$ the tip's roll length and $p_b=\pi m\cos\alpha$ the base pitch. Between
  them one tooth carries the whole load. Rating standards check bending at the HPSTC and contact stress at the pitch
  point or the LPSTC;
\item the \emph{pitch point}, $L_P=r\sin\alpha$, where sliding reverses;
\item the \emph{tip}, which is the end of active profile when there is no tip relief or chamfer.
\end{itemize}
Because $\inv\phi\approx\phi^3/3$ for small $\phi$, most of the involute's turn happens near the tip. In the worked
example the tooth thickness changes only from \thBase{} to \thSap~mm between the base circle and the SAP, but falls from
\thPitch~mm at the pitch circle to \thTip~mm at the tip. The author's notes draw three practical conclusions, which this
paper makes precise: compute in radians and in double precision; never compute $\tan\phi-\phi$ from angles rounded in
degrees; and space lookup tables of $\inv\phi$ logarithmically or more finely below about $10^\circ$.

\paragraph{Worked example.} Table~\ref{tab:example} gives the named diameters for $z=20$, $m=2$~mm, $\alpha=20^\circ$,
$x=0$, a rack cutter with addendum $h_{a0}=1.25m=2.50$~mm and tip radius $\rho_{a0}=0.38m=0.76$~mm (the ISO~53 profile~A
proportions~\cite{iso53}), and an identical mate at the standard centre distance $a=40$~mm. This gives $h'=\hPrime$~mm,
$L_{\rm TIF}=\lTif$~mm, $L_{\rm SAP}=\lSap$~mm, transverse contact ratio \exEps{} and base pitch \exPb~mm. A separate
simulation of the rack generation confirms that the flank is involute down to the TIF, and to within $0.001m$ even
slightly below it (\exTifNum~mm).

\begin{table}[ht]
\centering\small
\input{tables/gear_example.tex}
\caption{Worked example $z=20$, $m=2$~mm, $\alpha=20^\circ$, $x=0$, ISO~53~A rack, identical mate. $y=\inv\phi$
is the value the inverse involute must invert to recover $\phi$ at that diameter.}
\label{tab:example}
\end{table}

\paragraph{Undercut limits.} By this rack, gears with $x=0$ undercut for $z$ below \zminA{} at $14.5^\circ$, \zminB{} at
$20^\circ$, \zminC{} at $25^\circ$ and \zminD{} at $30^\circ$. The classical limit $2/\sin^2\alpha$ (\zClassA{} at
$14.5^\circ$, \zClassB{} at $20^\circ$) assumes a straight rack flank reaching $1.0m$ below the pitch line. With the
$0.38m$ tip radius (used up to $23.16^\circ$; Section~\ref{sec:gears}) the straight flank reaches $h'=\hPrimeA m$ at $14.5^\circ$ and exactly $1.0m$ at $20^\circ$, hence the
slightly lower limit at $14.5^\circ$. For $z=5$, $\alpha=14.5^\circ$ the end of the rack's straight flank lies $3.229m$ beyond the base-tangent point ($L_{\rm TIF}=\uGtif m$). At the minimum shift $x=\uXmin$ the computed tip thickness is
negative (\uShiftTip$m$). Over the envelope, \envShiftPointed{} of the \envUndercut{} undercut gears have pointed tips at their minimum shift. On these shifted or pointed teeth the local pressure angle reaches at most $55.85^\circ$ ($y=0.4995$, at the pointed tip of $z=5$, $\alpha=30^\circ$), inside the bound $y\le\envYShiftMax$ of Section~\ref{sec:gears}; for $z=5$, $\alpha=14.5^\circ$ the pointed tip lies at $\phi=\pointedPhiMax^\circ$ ($y=0.4033$).

\paragraph{SAP across the envelope.} With an identical mate at the standard centre distance, \sapNoValid{} of the gears
have no valid SAP (the mate's tip reaches past the base-tangent point, so the pair interferes without profile shift). If
undercut gears are included, the SAP can come arbitrarily close to the base circle: $\phi=\sapMinPhi^\circ$ at
$z=\sapMinZ$, $\alpha=\sapMinA^\circ$, which falls in the undercut region and is not a working flank point.

%% file: tables/gear_example.tex
\begin{tabular}{lrrrr}
\toprule
point & $d$ (mm) & $\phi$ ($^\circ$) & roll angle $\xi$ ($^\circ$) & $y=\inv\phi$ \\
\midrule
root & 35.000 & --- & --- & --- \\
base & 37.588 & --- & --- & --- \\
TIF & 37.640 & 3.024 & 3.027 & $4.908\times10^{-5}$ \\
SAP & 37.855 & 6.810 & 6.842 & $5.629\times10^{-4}$ \\
LPSTC & 39.182 & 16.402 & 16.866 & $8.085\times10^{-3}$ \\
pitch & 40.000 & 20.000 & 20.854 & 0.01490 \\
HPSTC & 40.969 & 23.441 & 24.842 & 0.02446 \\
tip & 44.000 & 31.321 & 34.866 & 0.06186 \\
\bottomrule
\end{tabular}

%% file: app_mfg.tex
\section{Manufacturing and CAD details}\label{app:mfg}

\paragraph{Hobs.} A hob is a worm whose axial section is the straight-sided basic rack. The involute is not on
the tool: it is produced by the rolling (generating) motion between hob and blank. Hob teeth are relieved (backed off)
so they can be resharpened, and the thread is ground on precision hobs. Errors in the hob's thread lead, pressure angle
or form carry into the gear as profile and base-pitch errors. A pressure-angle error $\delta\alpha$ changes the base
pitch by $\delta p_b=\pi m\sin\alpha\,\delta\alpha$, which is \hobPbTwo~$\mu$m per arc-minute at $m=2$~mm and \hobPbTen~$\mu$m
at $m=10$~mm. It tilts the profile by about $L_{AE}\tan\alpha\,\delta\alpha$, where $L_{AE}$ is the roll length of the active profile from the SAP to the tip, the span over which $f_{H\alpha}$ is extrapolated (Appendix~\ref{app:iso}): \hobFhTwo~$\mu$m per arc-minute for the worked example ($L_{AE}=\hobLaTwo$~mm; the 95\% evaluation range is $L_\alpha=\tbLa$~mm; class-5 $f_{H\alpha T}=\hobIsoTwo$~$\mu$m),
and \hobFhTen~$\mu$m for $m=10$, $z=100$ (class-5 $f_{H\alpha T}=\hobIsoTen$~$\mu$m).

\paragraph{Shaper cutters.} A shaper cutter is itself an involute gear. Its own profile accuracy transfers to the work,
and regrinding its face over its life changes its effective profile shift and outside diameter. This shifts the
generated root and TIF unless the setup is corrected.

\paragraph{Wire EDM from DXF.} Here the part is a one-to-one copy of the drawing, so the drawing must be right. A DXF
flank is a polyline, arcs or a spline, and CAM may convert it to G01/G02 moves. Points are often sampled by radius,
and the polar angle of each point is $\psi(R)$ from~\eqref{eq:halfangle}, that is, the forward involute. The radius
of curvature of the involute is the roll length $r_b\tan\phi$, smallest near the base circle, so the chord sag
$\approx s^2/(8\rho)$ is worst low on the flank. For the worked example, the flank from TIF to tip (\dxLen~mm long) needs
\dxPolyOne{} polyline points to keep the chord deviation within 1~$\mu$m (\dxPolyTen{} for 10~$\mu$m), or \dxArcOne{}
three-point arcs for 1~$\mu$m (\dxArcTenth{} for 0.1~$\mu$m). For $m=10$~mm, $z=100$ (\dxBigLen~mm of flank), the figures
are \dxBigPolyOne{} points or \dxBigArcOne{} arcs for 1~$\mu$m. Few points are needed, provided each comes from the true involute.

\paragraph{STEP in a CAD system.} In a STEP file the flank is usually a B-spline fitted to computed points, so the fit
tolerance and the accuracy of those points set the surface. A B-spline fitted to the true involute is good to the fit
tolerance. A curve that is not an involute stays wrong however finely it is fitted.

\paragraph{STEP to a 3D printer.} The STEP surface is tessellated to STL, sliced into layers and turned into toolpaths.
STL stores vertices in binary32, whose spacing is \stlSpacingSmall~$\mu$m at 22~mm and \stlSpacingBig~$\mu$m at 500~mm,
which is negligible. The printing process itself is far coarser than the arithmetic. But process errors scatter, while
a wrong source profile is a systematic error in every tooth. For printed gears that actually run, only the true involute
gives conjugate action, a constant ratio and tolerance to centre-distance change.

\begin{figure}[ht]
\centering
\includegraphics[width=0.9\textwidth]{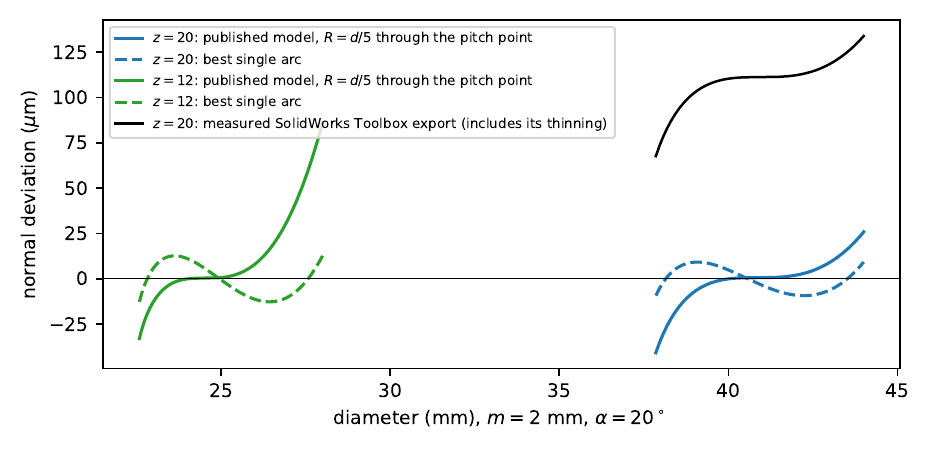}
\caption{Normal deviation of a circular-arc flank from the true involute, $m=2$~mm, $\alpha=20^\circ$. Coloured solid: the
published arc model of~\cite{ju_gears}. Black: the measured SolidWorks~2026 Toolbox export, $z=20$~\cite{toolbox2026}. Dashed: the best (minimax) single arc over the same range. Positive: the arc lies inside the
involute (tooth too thin).}
\label{fig:arc}
\end{figure}

\paragraph{Circular-arc teeth.} One published analysis~\cite{ju_gears} gives the SolidWorks Toolbox arc as radius $d/5$, centred on
the base circle and passing through the pitch point. We confirm its stated geometry for $z=12$, $m=6.35$~mm (arc centre
at $\arcGamma^\circ$ from the pitch-point radius). At the pitch point that arc is \arcFlat\% flatter than the involute,
whose radius of curvature there is $r\sin\alpha$. The SolidWorks~2026 Toolbox export~\cite{toolbox2026} (Section~\ref{sec:mfg})
has the same radius and centre circle but its centres are rotated \tbRot$^\circ$, which thins the tooth uniformly (the SolidWorks Toolbox sketch carries a backlash entry of \tbBL~mm, Section~\ref{sec:mfg}): its
deviation is the model curve shifted by about \tbNormOff~mm along the normal. Figure~\ref{fig:arc} shows the normal
deviation, $m=2$~mm, $\alpha=20^\circ$. For $z=20$, from the SAP ($d=\arcLo$~mm) to the tip, the published model runs
from \arcTbMin{} to $+\arcTbMax$~$\mu$m, the measured export from $+\tbDevMin$ to $+\tbDevMax$~$\mu$m, and the best single
arc stays within $\pm\arcBest$~$\mu$m. For $z=12$ (undercut at $x=0$), from the start of the remaining involute
($d=\arcLoB$~mm) to the tip, the model runs from \arcTbMinB{} to $+\arcTbMaxB$~$\mu$m, and the best arc stays within
$\pm\arcBestB$~$\mu$m. The best-arc and $z=12$ classes compare the span (maximum minus minimum) with $F_{\alpha T}$ over
the stated range. The measured $z=20$ export is evaluated as in ISO~1328-1:2013 (Appendix~\ref{app:iso}): $F_\alpha$,
$f_{f\alpha}$ and $f_{H\alpha}$ over $L_\alpha$, with plus material beyond the range near the tip included. For rigid gears a profile deviation along the normal is a displacement along
the line of action, so it appears directly as transmission error. Two arc teeth in contact behave as a four-bar
linkage, because the distance between their arc centres stays constant; the ratio quoted in Section~\ref{sec:mfg} is
over $\pm$ half a pitch about the pitch point and repeats every tooth.

\section{ISO 1328-1:2013 profile tolerances}\label{app:iso}
The class-5 values are multiplied by $\sqrt2^{A-5}$ for class $A$ (computed unrounded). The results are rounded to the
nearest 1~$\mu$m above 10~$\mu$m, to the nearest 0.5~$\mu$m from 5.0 to 10~$\mu$m, and to the nearest 0.1~$\mu$m below
5.0~$\mu$m; $F_{\alpha T}$ is formed from the unrounded $f_{H\alpha T}$ and $f_{f\alpha T}$. The standard does not say how
to break ties; we round halves up, which matters only for $f_{f\alpha T}=10.5$~$\mu$m at $m=10$, class~5.
Table~\ref{tab:iso} lists the values used, with the unrounded figures in parentheses.

\emph{Profile evaluation.} For the measured SolidWorks Toolbox flank (Section~\ref{sec:mfg}) we follow 4.4.8.2 and 3.4.1.4 of the
standard: the evaluation range $L_\alpha$ runs along the roll length from the SAP ($d_{Cf}$) and covers 95\% of the
length to the tip ($d_{Fa}$). The mean line is the least-squares line through the deviation over $L_\alpha$. $F_\alpha$ is
the band of the deviation between two design-profile lines, and $f_{f\alpha}$ is the band between two lines parallel
to the mean line. In the zone beyond $L_\alpha$ near the tip, plus material is included and minus material is ignored.
$f_{H\alpha}$ is the change of the mean line, extrapolated, from $d_{Cf}$ to the tip.

\begin{table}[ht]
\centering\small
\input{tables/gear_iso.tex}
\caption{ISO~1328-1:2013 profile tolerances ($\mu$m) for the gear sizes used in this paper, rounded per the standard
(unrounded in parentheses). Generated by the author's script.}
\label{tab:iso}
\end{table}

%% file: tables/gear_iso.tex
\begin{tabular}{llrrrrr}
\toprule
$m$, $d$ (mm) & & class 1 & class 2 & class 5 & class 8 & class 11 \\
\midrule
1, 20 & $f_{H\alpha T}$ & 1.1 (1.105) & 1.6 (1.563) & 4.4 (4.420) & 13 (12.502) & 35 (35.360) \\
 & $f_{f\alpha T}$ & 1.4 (1.387) & 2 (1.962) & 5.5 (5.550) & 16 (15.698) & 44 (44.400) \\
 & $F_{\alpha T}$ & 1.8 (1.774) & 2.5 (2.508) & 7 (7.095) & 20 (20.068) & 57 (56.760) \\
\addlinespace
2, 40 & $f_{H\alpha T}$ & 1.2 (1.210) & 1.7 (1.711) & 4.8 (4.840) & 14 (13.690) & 39 (38.720) \\
 & $f_{f\alpha T}$ & 1.5 (1.525) & 2.2 (2.157) & 6 (6.100) & 17 (17.253) & 49 (48.800) \\
 & $F_{\alpha T}$ & 1.9 (1.947) & 2.8 (2.753) & 8 (7.787) & 22 (22.025) & 62 (62.295) \\
\addlinespace
2, 80 & $f_{H\alpha T}$ & 1.2 (1.220) & 1.7 (1.725) & 4.9 (4.880) & 14 (13.803) & 39 (39.040) \\
 & $f_{f\alpha T}$ & 1.5 (1.525) & 2.2 (2.157) & 6 (6.100) & 17 (17.253) & 49 (48.800) \\
 & $F_{\alpha T}$ & 2 (1.953) & 2.8 (2.762) & 8 (7.812) & 22 (22.095) & 62 (62.494) \\
\addlinespace
10, 1000 & $f_{H\alpha T}$ & 2.2 (2.250) & 3.2 (3.182) & 9 (9.000) & 25 (25.456) & 72 (72.000) \\
 & $f_{f\alpha T}$ & 2.6 (2.625) & 3.7 (3.712) & 11 (10.500) & 30 (29.698) & 84 (84.000) \\
 & $F_{\alpha T}$ & 3.5 (3.457) & 4.9 (4.889) & 14 (13.829) & 39 (39.115) & 111 (110.635) \\
\addlinespace
\bottomrule
\end{tabular}

%% file: app_pins.tex
\section{Over-pins and over-balls study: details}\label{app:pins}
\paragraph{Assumptions.} Profile shift $x=0$; ISO~53~A rack ($1.25m_n$ dedendum; tip radius $0.38m_n$ up to $23.16^\circ$ and above that the largest that fits on the rack tooth, $0.32m_n$ at $25^\circ$ and $0.11m_n$ at $30^\circ$, as in Section~\ref{sec:gears}); addendum
$1m_n$; nominal tooth thickness $s_n=\pi m_n/2$ with no thickness tolerance unless stated. The root involute limit
$d_{\rm Lf}$ is the TIF from the rack. The tip involute limit $d_{\rm La}=d_a-0.1m_n$ assumes a $0.05m_n$ radial tip break,
which is our choice and not a value from the standard. Undercut gears are excluded, because their involute start needs the
generation simulation of Appendix~\ref{app:flank}. Helical gears use transverse geometry: the 9.5 relation for
$\inv\alpha_{kt}$ with $D_d/(z m_n\cos\alpha_n)$, contact by its Eq.~(98) with the factor $\cos\beta_b$, and two balls in a
transverse plane for odd $z$ (9.7.2). The pin for a desired contact diameter is found by solving those two relations,
which is equivalent to the iteration of 9.4.8. Internal gears are treated in
Appendix~\ref{app:pinsint}.

\paragraph{Envelope results.} Over the \pinGears{} gears: with contact at the TIF, $\alpha_k$ ranges from \pinLowAkMin$^\circ$
to \pinLowAkMax$^\circ$ and the gain from \pinLowGainMin{} to \pinLowGainMax. No pin sized for TIF contact projects past
the tip (\pinLowProjPct\%), and \pinLowRootPct\% of them would touch the root. The lowest projecting pin has
$\alpha_k\ge\pinProjAkMin^\circ$ and gain at most \pinProjGainMax. At the midpoint the gain ranges from \pinMidGain{} to
\pinMidGainMax, and at $d_{\rm La}$ from \pinHighGainMin{} to \pinHighGainMax, with $y$ up to \pinHighYMax{}
($\alpha_k$ up to \pinHighAkMax$^\circ$), still well inside the proved range $y<2$. The $1.728m_n$ pin gives contact on the
involute, clear of the root and projecting past the tip, for every gear in the envelope, below the midpoint for
\pinStdBelowPct\% of them. The inverse-method errors were evaluated against 80-digit roots for every $z$ at
$\alpha=14.5^\circ$, $20^\circ$, $25^\circ$ and $30^\circ$ (\pinNErr{} gears per position). Cheng's largest errors per
millimetre of module are \pinChengLowMax~$\mu$m at the TIF, \pinChengMidMax~$\mu$m at the midpoint (both at $z=500$,
$\alpha=30^\circ$) and \pinChengHighMax~$\mu$m at $d_{\rm La}$.

\begin{figure}[ht]
\centering
\includegraphics[width=\textwidth]{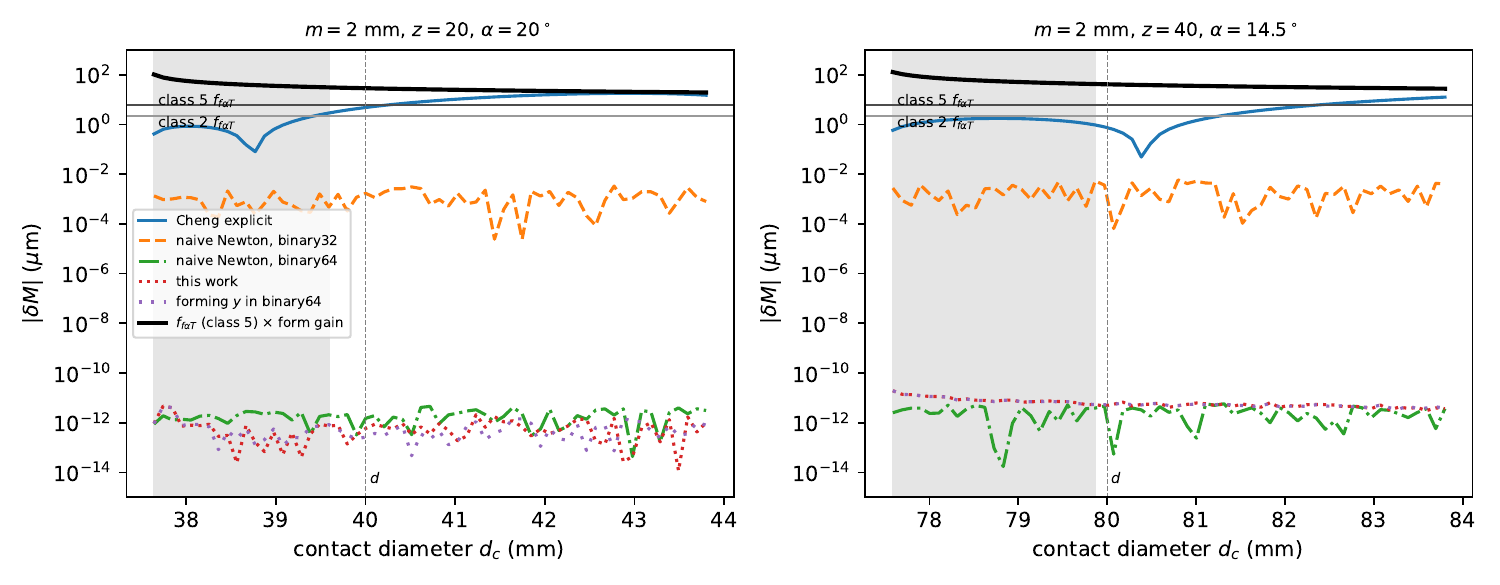}
\caption{Error in the measurement over pins against contact diameter, from $d_{\rm Lf}$ to $d_{\rm La}$, for each way of
computing the inverse. The heavy line is the effect of a local form deviation equal to the class-5 $f_{f\alpha T}$; grey
horizontal lines are the class-2 and class-5 $f_{f\alpha T}$. In the shaded band the pin does not project past the tip.
Dashed vertical line: reference diameter.}
\label{fig:pins}
\end{figure}

\subsection{Internal gears: measurement between pins and balls}\label{app:pinsint}
ANSI/AGMA 2002-D19 covers internal gears with a negative tooth count (its 4.7.1), as ISO~21771-1:2024 does (its 4.2). For
an internal gear the measurement is \emph{between} the pins or balls, and its 9.5 and 9.7 relations become, for $|z|=N$,
\begin{equation}
y=\inv\alpha_k=\inv\alpha_t+\frac{\pi-s_n/m_n}{N}-\frac{D_d}{N m_n\cos\alpha_n},\qquad
\tan\alpha_c=\tan\alpha_k+\frac{D_d\cos\beta_b}{d_b},
\label{eq:pinsint}
\end{equation}
and $M=k\,d_b/\cos\alpha_k-D_d$, with diameters taken as positive magnitudes and $k=\cos(\pi/2N)$ for odd $N$ and $k=1$ otherwise. Here $\alpha_c$ is the transverse pressure angle at the contact
point. The inverse is again taken at the pin-centre angle $\alpha_k$. The sign of the contact term is reversed, however.
On an external gear the pin centre lies further out along the line of action than the contact point, so
$\alpha_k>\alpha_c$. On an internal gear it lies nearer the base circle, so $\alpha_k<\alpha_c$. A tooth-thickness change
still moves $M$ by $|\partial M/\partial s_n|=k\cos\alpha_t/(\cos\beta\sin\alpha_k)$, and the form gain is as before.
Pins are used only for spur gears and balls for helical gears (9.1, 9.2).

\paragraph{Model and envelope.} These choices are ours. Profile shift is $x=0$ and $s_n=\pi m_n/2$ (the internal
\emph{tooth}, 4.7.1). The internal tooth has a $1m_n$ addendum, so the minor (tip) radius is $r-m_n$, and a $1.25m_n$
dedendum. The gear is generated by a pinion-type cutter with $x_0=0$, a $1.25m_n$ addendum, and a corner radius of
$0.38m_n$ (less if the cutter tooth is too narrow for it, and at least $0.05m_n$). The root involute limit $d_{\rm Lf}$ is
the root form diameter of ISO~21771-1:2024 Formulas (259)--(262) for that cutter. The tip involute limit sits $0.05m_n$
radially outside the minor circle, which is the internal equivalent of the $0.05m_n$ tip break used for external gears.
The mating pinion has the cutter's tooth count $z_0$ and a $1m_n$ addendum. For each gear $(N,\alpha)$ we take the
smallest $z_0\ge\max(25,\lceil 2/\sin^2\alpha\rceil)$ that passes three checks, and the gear is in the envelope only if
such a $z_0<N$ exists. First, its minor circle lies outside the base circle, so the whole flank is involute. Second,
the pinion does not interfere with it in the sense of ISO~21771-1:2024 5.5.8.2, that is, $CA<CT_1$ and $d_{\rm Nf2}<d_{\rm
Ff2}$. Third, neither the pinion nor the cutter meets it tip to tip by the test of 5.5.8.3, Formulas (105)--(110), which we
also use as the overcut check that 9.9 suggests. We ignored the cutter's radial infeed stroke.

Over $14.5^\circ$ to $30^\circ$ in $0.5^\circ$ steps and $N\le500$, \intGears{} of the \intGrid{} candidate gears (every $N$ from 10 to 500 at each of the 32 angles) qualify.
The smallest qualifying tooth count runs from \intNminLo{} (at $28^\circ$) to \intNminHi{} (at $30^\circ$): $N\ge\intNminA$ at $14.5^\circ$ (with a
\intZoA-tooth pinion), $N\ge\intNminB$ at $20^\circ$ (\intZoB-tooth pinion), $N\ge\intNminC$ at $25^\circ$, and $N\ge\intNminD$
at $30^\circ$. At the smallest qualifying count, the gear has between \intDzLo{} and \intDzHi{} more teeth than its pinion.
The $30^\circ$ limit comes from the cutter rather than the gear. A $1.25m_n$-deep cutter tooth at $30^\circ$ is nearly
pointed, and only a \intZoD-tooth cutter leaves room for even a $0.05m_n$ corner radius. A shallower or shifted tool would
move this limit. The low local pressure angles are now at the \emph{tip} of the internal tooth, on the minor diameter
next to the base circle. At the root, the flank meets the fillet at a local pressure angle above the nominal one.

\begin{figure}[ht]
\centering
\includegraphics[width=\textwidth]{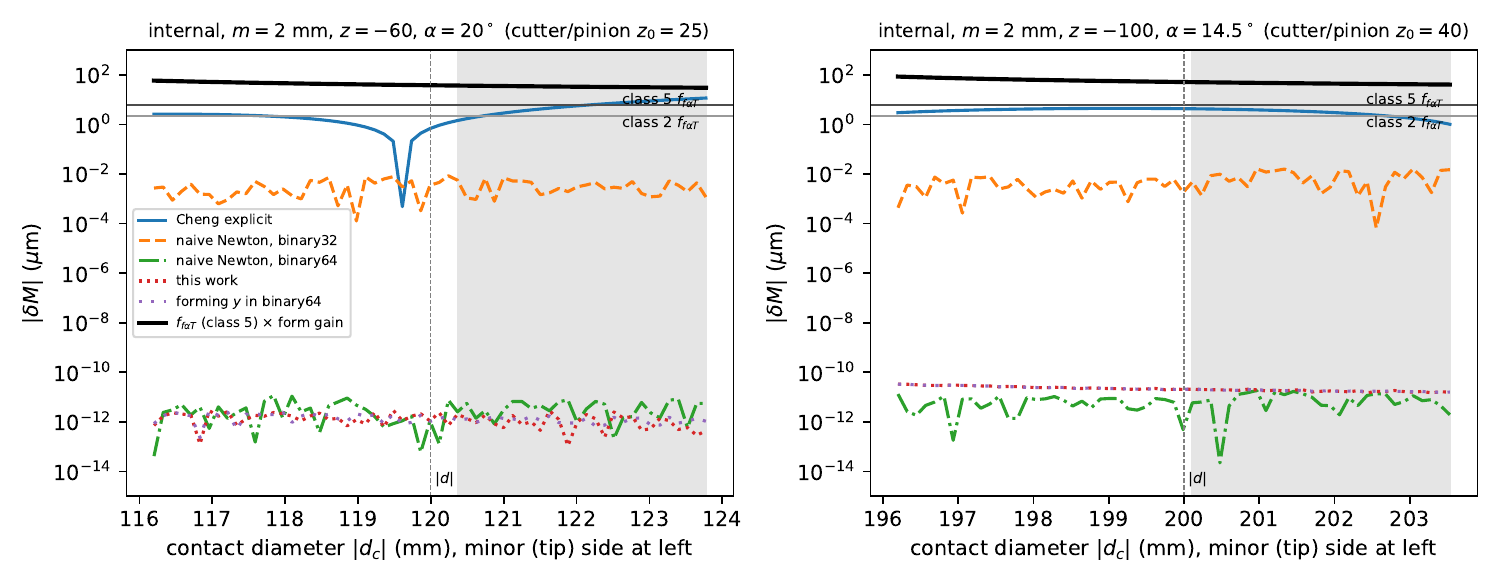}
\caption{Internal gears: error in the measurement between pins against contact diameter, from the minor-side limit
$d_{\rm La}$ (left) to the root form diameter $d_{\rm Lf}$ (right). Lines and shading as in Figure~\ref{fig:pins}. In the
shaded band the pin does not project inside the minor circle.}
\label{fig:pinsint}
\end{figure}

\paragraph{Results.} We took seven contact positions per gear: at $d_{\rm La}$ (minor side), at the last contact whose pin
still projects inside the minor circle, at the 9.4.8 midpoint and at the reference diameter, with the $1.68m_n$ and
$1.44m_n$ internal pins of 9.4.8, and at $d_{\rm Lf}$. Table~\ref{tab:pinsint} and Figure~\ref{fig:pinsint} show
representative gears. Over the envelope:
\begin{itemize}[nosep]
\item \emph{Inputs.} At the minor-side limit, $\alpha_k$ ranges from \intTipAkMin$^\circ$ to \intTipAkMax$^\circ$ and $y$ from
  \intTipYMin{} to \intTipYMax. Here $\alpha_k$ is \intTipLagMin$^\circ$ to \intTipLagMax$^\circ$ \emph{below} the local
  pressure angle at the contact. At the root limit, $\alpha_k$ reaches \intRootAkMax$^\circ$ ($y\le\intRootYMax$). All of these
  inputs lie inside the range already exercised by the external gears ($y\ge\extLowYMin$, $\alpha_k\ge\extLowAkMin^\circ$ at
  their TIF) and inside the proved range of the inverse.
\item \emph{Arithmetic.} At every position, the errors in $M$ from this work, from naive binary64 Newton and from forming
  $y$ in binary64 stay below \intConvMax~$\mu$m per millimetre of module, and binary32 Newton stays below
  \intBtwoMax~$\mu$m. Cheng's formula behaves as it does on external gears. Its error tends to be smallest at the low-angle (minor) end and largest at the root, up to \intTipCheng, \intMidCheng{} and \intRootCheng~$\mu$m per millimetre of module
  at $d_{\rm La}$, at the midpoint and at $d_{\rm Lf}$ ($N=\intChengZ$, $\alpha=\intChengA^\circ$), against
  \pinChengLowMax--\pinChengHighMax~$\mu$m for external gears. These were evaluated against 80-digit roots for every $N$
  at $14.5^\circ$, $20^\circ$, $25^\circ$ and $30^\circ$ (\intNErr{} gears per position).
\item \emph{Physical amplification.} The form gain is largest at the minor-side limit. It reaches \intTipGainMax{} (at
  $N=\intTipGainZ$, $\alpha=\intTipGainA^\circ$), against \intMidGainMax{} at the midpoint and \intRootGainMax{} at the root.
  The contact also moves by up to \intTipDrc{} times the thickness change there, against \intMidDrc{} at the midpoint. On
  the $m=2$, $z=-60$, $20^\circ$ gear, a local deviation equal to the class-5 $f_{f\alpha T}$ moves $M$ by about
  \exiTipFf~$\mu$m at the minor-side limit and \exiMidFf~$\mu$m at the midpoint. Because $\alpha_k<\alpha_c$ on an internal
  gear, the gain is somewhat larger than the contact angle alone would suggest.
\item \emph{Measurability.} The two ends swap roles. The largest pins contact at the minor side, and every such pin
  projects inside the minor circle and clears the root (\intTipProjPct\%). No pin sized for root-limit contact projects
  (\intRootProjPct\%), and \intRootTouchPct\% of them would touch the root. The midpoint pin projects in \intMidProjPct\% of
  the gears. The $1.68m_n$ pin contacts the involute, clears the root and projects for every gear, on the minor side of the
  midpoint in \intPbTipsidePct\% of them. The $1.44m_n$ pin contacts the involute everywhere but never projects inside the
  minor circle with these proportions. It suits only instruments that do not need projection (9.4.5).
\end{itemize}
Internal gears add no new numerical difficulty. The inputs stay in the range the external study covers, and a
converged inverse contributes nothing measurable. The new effects are physical. The low-angle, high-gain end of an internal
flank is the end where pins are easiest to fit, so a pin chosen for convenience can land where form error is amplified
most. The midpoint and the $1.68m_n$ pin remain the well-conditioned choices.

\begin{table}[ht]
\centering\scriptsize\setlength{\tabcolsep}{2.5pt}
\resizebox{\textwidth}{!}{\input{tables/pins_internal.tex}}
\caption{Internal gears: measurement between two pins or balls against contact position (ANSI/AGMA 2002-D19 rules with
$z<0$; nominal tooth thickness; our cutter and pinion model). $\alpha_c$: transverse pressure angle at the contact;
$\alpha_k$: at the pin centre, where the inverse is evaluated. Proj.: the pin projects inside the minor circle (9.4.5);
root: it clears the major (root) circle (9.4.6). Other columns as in Table~\ref{tab:pinsloc}. Generated by the author's script.}
\label{tab:pinsint}
\end{table}

%% file: tables/pins_internal.tex
\begin{tabular}{lrrrrrrrrrrcc}
\toprule
contact & $|d_c|$ & $D_d$ & $\alpha_c$ & $\alpha_k$ & $y$ & gain & gain$\cdot f_{f\alpha T}^{(5)}$ & Cheng & b32 & this work & proj. & root \\
 & (mm) & (mm) & ($^\circ$) & ($^\circ$) & & & ($\mu$m) & ($\mu$m) & ($\mu$m) & ($\mu$m) & & \\
\midrule
\multicolumn{13}{l}{\emph{internal spur, $m=2$, $z=-60$, $\alpha=20^\circ$ (pins)}; cutter/pinion $z_0=25$; $f_{f\alpha T}$ class 2/5/8: 2.2/6/17~$\mu$m} \\
$d_{\rm La}$ (minor side) & 116.20 & 4.290 & 14.0 & 11.9 & 3.04e-3 & 9.7 & 58 & 2.6 & $2.7\times10^{-3}$ & $8.6\times10^{-13}$ & \checkmark & \checkmark \\
last proj. & 120.30 & 3.214 & 20.4 & 18.9 & 1.26e-2 & 6.2 & 37 & 1.3 & $5.6\times10^{-4}$ & $2.0\times10^{-12}$ & \checkmark & \checkmark \\
midpoint & 119.99 & 3.316 & 20.0 & 18.5 & 1.17e-2 & 6.3 & 38 & 0.68 & $3.5\times10^{-3}$ & $1.1\times10^{-12}$ & \checkmark & \checkmark \\
$1.68m_n$ & 119.85 & 3.360 & 19.8 & 18.3 & 1.13e-2 & 6.4 & 38 & 0.43 & $4.6\times10^{-3}$ & $1.9\times10^{-12}$ & \checkmark & \checkmark \\
$1.44m_n$ & 121.24 & 2.880 & 21.5 & 20.3 & 1.55e-2 & 5.8 & 35 & 3.5 & $4.2\times10^{-3}$ & $1.8\times10^{-12}$ & -- & \checkmark \\
$d_{\rm Lf}$ (root form) & 123.78 & 1.823 & 24.4 & 23.6 & 2.49e-2 & 5.0 & 30 & 12 & $1.2\times10^{-3}$ & $1.1\times10^{-12}$ & -- & \checkmark \\
\addlinespace
\multicolumn{13}{l}{\emph{internal spur, $m=2$, $z=-37$ (smallest in the envelope at $20^\circ$), $\alpha=20^\circ$ (pins)}; cutter/pinion $z_0=26$; $f_{f\alpha T}$ class 2/5/8: 2.2/6/17~$\mu$m} \\
$d_{\rm La}$ (minor side) & 70.20 & 3.976 & 7.9 & 4.6 & 1.78e-4 & 24.7 & 148 & 0.41 & $3.0\times10^{-3}$ & $7.2\times10^{-13}$ & \checkmark & \checkmark \\
last proj. & 74.28 & 3.211 & 20.6 & 18.2 & 1.12e-2 & 6.4 & 38 & 0.22 & $1.4\times10^{-3}$ & $2.8\times10^{-13}$ & \checkmark & \checkmark \\
midpoint & 74.10 & 3.265 & 20.2 & 17.8 & 1.04e-2 & 6.5 & 39 & 0.071 & $4.5\times10^{-3}$ & $1.2\times10^{-12}$ & \checkmark & \checkmark \\
$1.68m_n$ & 73.77 & 3.360 & 19.5 & 17.0 & 9.04e-3 & 6.8 & 41 & 0.55 & $3.1\times10^{-3}$ & $9.3\times10^{-13}$ & \checkmark & \checkmark \\
$1.44m_n$ & 75.26 & 2.880 & 22.5 & 20.4 & 1.59e-2 & 5.7 & 34 & 2.3 & $1.1\times10^{-3}$ & $6.1\times10^{-13}$ & -- & \checkmark \\
$d_{\rm Lf}$ (root form) & 78.00 & 1.678 & 26.9 & 25.8 & 3.32e-2 & 4.6 & 28 & 12 & $2.3\times10^{-3}$ & $4.1\times10^{-13}$ & -- & \checkmark \\
\addlinespace
\multicolumn{13}{l}{\emph{internal spur, $m=2$, $z=-100$, $\alpha=14.5^\circ$ (pins)}; cutter/pinion $z_0=40$; $f_{f\alpha T}$ class 2/5/8: 2.2/6/17~$\mu$m} \\
$d_{\rm La}$ (minor side) & 196.20 & 3.928 & 9.3 & 8.1 & 9.67e-4 & 14.1 & 85 & 3.0 & $4.3\times10^{-4}$ & $3.3\times10^{-11}$ & \checkmark & \checkmark \\
last proj. & 200.01 & 3.229 & 14.5 & 13.6 & 4.58e-3 & 8.5 & 51 & 4.3 & $7.0\times10^{-3}$ & $2.1\times10^{-11}$ & \checkmark & \checkmark \\
midpoint & 199.87 & 3.263 & 14.4 & 13.4 & 4.40e-3 & 8.6 & 52 & 4.3 & $5.9\times10^{-3}$ & $2.1\times10^{-11}$ & \checkmark & \checkmark \\
$1.68m_n$ & 199.44 & 3.360 & 13.9 & 12.9 & 3.90e-3 & 8.9 & 54 & 4.4 & $2.9\times10^{-3}$ & $2.1\times10^{-11}$ & \checkmark & \checkmark \\
$1.44m_n$ & 201.40 & 2.880 & 16.0 & 15.2 & 6.38e-3 & 7.6 & 46 & 3.5 & $9.3\times10^{-3}$ & $2.0\times10^{-11}$ & -- & \checkmark \\
$d_{\rm Lf}$ (root form) & 203.53 & 2.258 & 17.9 & 17.3 & 9.59e-3 & 6.7 & 40 & 1.0 & 0.015 & $1.6\times10^{-11}$ & -- & -- \\
\addlinespace
\multicolumn{13}{l}{\emph{internal spur, $m=10$, $z=-200$, $\alpha=20^\circ$ (pins)}; cutter/pinion $z_0=25$; $f_{f\alpha T}$ class 2/5/8: 3.7/11/30~$\mu$m} \\
$d_{\rm La}$ (minor side) & 1981.00 & 23.266 & 18.4 & 17.8 & 1.04e-2 & 6.5 & 72 & 2.2 & 0.039 & $4.2\times10^{-11}$ & \checkmark & \checkmark \\
last proj. & 2001.57 & 16.089 & 20.1 & 19.7 & 1.42e-2 & 5.9 & 65 & 41 & 0.065 & $1.9\times10^{-11}$ & \checkmark & \checkmark \\
midpoint & 1999.32 & 16.919 & 19.9 & 19.5 & 1.38e-2 & 6.0 & 66 & 35 & 0.016 & $1.5\times10^{-11}$ & \checkmark & \checkmark \\
$1.68m_n$ & 1999.64 & 16.800 & 20.0 & 19.5 & 1.38e-2 & 6.0 & 66 & 36 & 0.091 & $3.5\times10^{-11}$ & \checkmark & \checkmark \\
$1.44m_n$ & 2006.04 & 14.400 & 20.5 & 20.1 & 1.51e-2 & 5.8 & 64 & 52 & 0.057 & $1.6\times10^{-11}$ & -- & \checkmark \\
$d_{\rm Lf}$ (root form) & 2017.64 & 9.817 & 21.3 & 21.1 & 1.75e-2 & 5.6 & 61 & 84 & $7.9\times10^{-3}$ & $1.4\times10^{-11}$ & -- & \checkmark \\
\addlinespace
\multicolumn{13}{l}{\emph{internal helical, $m_n=2$, $z=-80$, $\alpha_n=20^\circ$, $\beta=15^\circ$ (balls)}; cutter/pinion $z_0=25$; $f_{f\alpha T}$ class 2/5/8: 2.2/6/17~$\mu$m} \\
$d_{\rm La}$ (minor side) & 161.84 & 4.455 & 16.7 & 15.2 & 6.46e-3 & 7.8 & 47 & 2.8 & $2.0\times10^{-3}$ & $5.1\times10^{-12}$ & \checkmark & \checkmark \\
last proj. & 165.95 & 3.216 & 20.9 & 19.9 & 1.47e-2 & 6.1 & 36 & 3.9 & $1.2\times10^{-4}$ & $2.0\times10^{-12}$ & \checkmark & \checkmark \\
midpoint & 165.60 & 3.337 & 20.6 & 19.6 & 1.39e-2 & 6.2 & 37 & 3.0 & 0.011 & $2.5\times10^{-12}$ & \checkmark & \checkmark \\
$1.68m_n$ & 165.53 & 3.360 & 20.5 & 19.5 & 1.37e-2 & 6.2 & 37 & 2.9 & $4.8\times10^{-3}$ & $2.9\times10^{-12}$ & \checkmark & \checkmark \\
$1.44m_n$ & 166.87 & 2.880 & 21.7 & 20.8 & 1.69e-2 & 5.8 & 35 & 6.3 & $4.0\times10^{-3}$ & $1.9\times10^{-12}$ & -- & \checkmark \\
$d_{\rm Lf}$ (root form) & 169.36 & 1.866 & 23.8 & 23.2 & 2.37e-2 & 5.2 & 31 & 14 & $2.8\times10^{-3}$ & $1.8\times10^{-12}$ & -- & \checkmark \\
\addlinespace
\bottomrule
\end{tabular}

%% file: tables/gear_forward.tex
\begin{tabular}{lrrrr}
\toprule
case & $d_y$ (mm) & \multicolumn{3}{c}{error in $s_y$ ($\mu$m): binary64 / binary32 / binary16} \\
\midrule
$m=2$, $z=20$, SAP & 37.85 & $1.1\times10^{-13}$ & $1.2\times10^{-4}$ & 0.51 \\
$m=10$, $z=100$, $\phi=5^\circ$ & 943.28 & $6.3\times10^{-12}$ & $2.1\times10^{-3}$ & 21 \\
$m=10$, $z=100$, $\phi=1^\circ$ & 939.84 & $3.8\times10^{-14}$ & $7.8\times10^{-4}$ & 1.7 \\
\bottomrule
\end{tabular}

%% file: tables/reference80.tex
\item[$10^{-3}$] \texttt{\seqsplit{3.3333346666672065575920000755167242436387291772156146033608014332136219530560943e-10}}
\item[$10^{-2}$] \texttt{\seqsplit{3.333466672063710975421715492830118939486219789889269803256519459581531016523127e-7}}
\item[$10^{-1}$] \texttt{\seqsplit{3.3467208545054511396337942223222500966424183444922383976653426574660632039916469e-4}}
\item[$20^\circ$ (0.349)] \texttt{\seqsplit{1.4904383867336443346617591069851471577984306150786772325745673649157149943593543e-2}}
\item[$1.0$] \texttt{\seqsplit{5.5740772465490223050697480745836017308725077238152003838394660569886139715172729e-1}}

%% file: main.bbl
\begin{thebibliography}{40}
\providecommand{\natexlab}[1]{#1}
\providecommand{\url}[1]{\texttt{#1}}
\expandafter\ifx\csname urlstyle\endcsname\relax
  \providecommand{\doi}[1]{doi: #1}\else
  \providecommand{\doi}{doi: \begingroup \urlstyle{rm}\Url}\fi

\bibitem[{AGMA}(2002)]{agma2015}
{AGMA}.
\newblock {{ANSI/AGMA} 2015-1-A01: Accuracy Classification System ---
  Tangential Measurements for Cylindrical Gears}.
\newblock American Gear Manufacturers Association, Alexandria, VA, 2002.
\newblock Approved by the AGMA membership June 2001 and as an American National
  Standard 1~August 2002; reaffirmed 2014 (R2014).

\bibitem[{AGMA}(2019)]{agma2002}
{AGMA}.
\newblock {{ANSI/AGMA} 2002-D19: Tooth Thickness and Backlash Measurement of
  Cylindrical Involute Gearing}.
\newblock American Gear Manufacturers Association, Alexandria, VA, 2019.
\newblock Revision of ANSI/AGMA 2002-C16; reaffirmed December 2, 2024.

\bibitem[Barbosa et~al.(2022)]{cvc5}
Haniel Barbosa et~al.
\newblock cvc5: A versatile and industrial-strength {SMT} solver.
\newblock In \emph{Tools and Algorithms for the Construction and Analysis of
  Systems (TACAS)}, volume 13243 of \emph{LNCS}, pages 415--442, 2022.

\bibitem[{Brown \& Sharpe Manufacturing Company}(1886)]{brownsharpe1886}
{Brown \& Sharpe Manufacturing Company}.
\newblock \emph{Practical Treatise on Gearing}.
\newblock Brown \& Sharpe Mfg. Co., Providence, RI, 1886.
\newblock Ch.~IV: lines of pressure at $75\tfrac12^\circ$ to the radius;
  Ch.~XI: rack interference from 31 teeth.

\bibitem[Camus(1735)]{camus1733}
Charles-\'Etienne-Louis Camus.
\newblock Sur la figure des dents des rou\"es, et des a\^iles des pignons, pour
  rendre les horloges plus parfaites.
\newblock \emph{Histoire de l'Acad\'emie royale des sciences, ann\'ee 1733,
  avec les M\'emoires de math\'ematique et de physique}, pages 117--140, 1735.
\newblock Memoir for the year 1733, printed 1735.

\bibitem[Camus(1806)]{camus1806}
Charles-\'Etienne-Louis Camus.
\newblock \emph{A Treatise on the Teeth of Wheels, Pinions, \&c., Demonstrating
  the Best Forms which can be Given to Them for the Various Purposes of
  Machinery}.
\newblock J. Taylor, London, 1806.
\newblock Translated from the French by J.~I. Hawkins; a translation of books
  10 and 11 of Camus's \emph{Cours de math\'ematique} (third part, 1751--1752).

\bibitem[Cheng(1992)]{cheng1992}
Harry~H. Cheng.
\newblock Derivation of the explicit solution of the inverse involute function
  and its applications.
\newblock In \emph{Proc. 6th ASME International Power Transmission and Gearing
  Conference}, Phoenix, AZ, 1992.

\bibitem[Coy(1983)]{coy1983}
John~J. Coy.
\newblock Geared power transmission technology.
\newblock In \emph{Advanced Power Transmission Technology}, pages 49--77,
  Cleveland, OH, 1983. NASA Lewis Research Center.
\newblock NTRS 19830011851; p.~52: de La Hire ``is said to have \ldots{}
  recommended the involute curve for gear teeth (ca.~1694)''.

\bibitem[Crosher(2013)]{crosher2013}
William~P. Crosher.
\newblock {Cycloidal teeth must not only be accurately spaced and shaped; their
  wheel centers must be fixed with equal care}.
\newblock Gear Solutions, Tooth Tips column,
  \url{https://gearsolutions.com/departments/tooth-tips-william-crosher-78/},
  2013.
\newblock Published 3 May 2013 (May 2013 issue). Quotes Wilfred Lewis, dated
  there to 1900. Accessed 2026-11-21.

\bibitem[{Dassault Syst\`emes}(2021)]{swtoolbox}
{Dassault Syst\`emes}.
\newblock {{SOLIDWORKS} Help: Gears (Toolbox)}.
\newblock
  \url{https://my.solidworks.com/reader/onlinehelp/2021%252Fenglish%252Fsolidworks%252Ftoolbox%252Fc_gears.htm/gears},
  2021.
\newblock 2021 help: ``Toolbox gears are representations for machine design
  purposes. They are not true involute gears that you can use for
  manufacturing.'' Accessed 2026-11-21.

\bibitem[{Dassault Syst\`emes}(2026)]{toolbox2026}
{Dassault Syst\`emes}.
\newblock {{SolidWorks} 2026 Toolbox, ISO spur gear $m=2$~mm, $z=20$,
  $\alpha=20^\circ$, face width 12~mm}.
\newblock STEP AP214 export, 2026.
\newblock Part generated and exported by the author on 2026-11-21 and measured
  for this paper. The CAD files and the derived geometry are not distributed.

\bibitem[de~Dinechin et~al.(2011)de~Dinechin, Lauter, and
  Melquiond]{dedinechin2011}
Florent de~Dinechin, Christoph Lauter, and Guillaume Melquiond.
\newblock Certifying the floating-point implementation of an elementary
  function using {G}appa.
\newblock \emph{IEEE Transactions on Computers}, 60\penalty0 (2):\penalty0
  242--253, 2011.

\bibitem[de~La~Hire(1694)]{delahire1694}
Philippe de~La~Hire.
\newblock Trait\'e des \'epicyclo\"ides et de leur usage dans les
  m\'echaniques.
\newblock Paris, 1694.

\bibitem[de~Moura and Bj{\o}rner(2008)]{z3}
Leonardo de~Moura and Nikolaj Bj{\o}rner.
\newblock {Z3}: An efficient {SMT} solver.
\newblock In \emph{Tools and Algorithms for the Construction and Analysis of
  Systems (TACAS)}, volume 4963 of \emph{LNCS}, pages 337--340, 2008.

\bibitem[{Deutsches Institut f\"ur Normung}(1927)]{din867}
{Deutsches Institut f\"ur Normung}.
\newblock {{DIN} 867:1927-07, Zahnform f\"ur Stirnr\"ader und Kegelr\"ader}.
\newblock Beuth/DIN Media catalogue record,
  \url{https://www.dinmedia.de/de/norm/din-867/111764524}, 1927.
\newblock First edition, July 1927; withdrawn. Later editions 1963-09, 1974-09,
  1984-05 and 1986-02 (current; DIN Media,
  \url{https://www.dinmedia.de/de/norm/din-867/1275267}). Accessed 2026-11-21.

\bibitem[{Deutsches Institut f\"ur Normung}(1974)]{din867later}
{Deutsches Institut f\"ur Normung}.
\newblock {{DIN} 867:1974-09, Bezugsprofil f\"ur Stirnr\"ader mit
  Evolventenverzahnung (basic rack tooth profile for involute cylindrical
  gears)}.
\newblock Beuth/DIN English translation, 2~pp., copy held by the author,
  September 1974.
\newblock Clause~4 and its figure give $\alpha_P=20^\circ$. Replaces DIN 867,
  Issue of September 1963.

\bibitem[Dupont(1963)]{dupont1963}
Pascal Dupont.
\newblock Nuove dimostrazioni della formula di {Eulero--Savary} e suo studio
  storica.
\newblock \emph{Atti della Accademia delle Scienze di Torino, Classe di Scienze
  Fisiche, Matematiche e Naturali}, 97:\penalty0 927--962, 1963.
\newblock Volume for 1962/1963; title as given by
  MacTutor~\cite{mactutorsavary}. Not seen by us.

\bibitem[{Ecma International}(2025)]{ecma262}
{Ecma International}.
\newblock {{ECMA-262}: {ECMAScript} Language Specification}.
\newblock \url{https://tc39.es/ecma262/}, 2025.
\newblock Number type: IEEE 754-2019 binary64, round to nearest, ties to even.

\bibitem[Euler(1760)]{euler1760}
Leonhard Euler.
\newblock De aptissima figura rotarum dentibus tribuenda.
\newblock \emph{Novi Commentarii Academiae Scientiarum Imperialis
  Petropolitanae}, 5 (for 1754--1755):\penalty0 299--316, 1760.
\newblock Enestr\"om index E249; \emph{Opera Omnia} II.17, 119--135. Catalogue
  and scan: \url{https://scholarlycommons.pacific.edu/euler-works/249/}.

\bibitem[Euler(1767)]{euler1767}
Leonhard Euler.
\newblock Supplementum de figura dentium rotarum.
\newblock \emph{Novi Commentarii Academiae Scientiarum Imperialis
  Petropolitanae}, 11 (for 1765):\penalty0 207--231, 1767.
\newblock Enestr\"om index E330, catalogued by the Euler Archive as \emph{De
  figura dentium rotarum}; \emph{Opera Omnia} II.17, 196--219; read to the
  Berlin Academy on 14 October 1762 according to the Euler Archive,
  \url{https://scholarlycommons.pacific.edu/euler-works/330/}. Section and page
  numbers cited here are from the Euler Archive scan of the original, which
  lacks p.~215.

\bibitem[Gladman et~al.(2024)Gladman, Innocente, Mather, and
  Zimmermann]{gladman2024}
Brian Gladman, Vincenzo Innocente, John Mather, and Paul Zimmermann.
\newblock Accuracy of mathematical functions in single, double, double
  extended, and quadruple precision.
\newblock Inria, HAL preprint hal-03141101, version~7,
  \url{https://inria.hal.science/hal-03141101v7}, August 2024.
\newblock Version of 7 August 2024; the report is updated regularly (version~8,
  17 February 2025, is the latest at the time of writing).

\bibitem[{Hawk Ridge Systems}(2024)]{hawkridge2024}
{Hawk Ridge Systems}.
\newblock {How to Determine if the {SOLIDWORKS} Toolbox Spur Gear Component is
  a Good Fit}.
\newblock
  \url{https://hawkridgesys.com/blog/solidworks-toolbox-spur-gear-component-is-a-good-fit},
  2024.
\newblock Accessed 2026-11-21.

\bibitem[{IEEE}(2019)]{ieee754}
{IEEE}.
\newblock {{IEEE} Standard for Floating-Point Arithmetic ({IEEE} Std
  754-2019)}.
\newblock IEEE, 2019.

\bibitem[{ISO}(1998)]{iso53}
{ISO}.
\newblock {{ISO} 53: Cylindrical gears for general and heavy engineering ---
  Standard basic rack tooth profile}.
\newblock International Organization for Standardization, Geneva, 1998.
\newblock Second edition, 1998-08, replacing ISO 53:1974; confirmed 2026 and
  current according to \url{https://www.iso.org/standard/22643.html} (accessed
  2026-11-21).

\bibitem[{ISO}(2013)]{iso1328}
{ISO}.
\newblock {{ISO} 1328-1:2013, Cylindrical gears --- {ISO} system of flank
  tolerance classification --- Part 1: Definitions and allowable values of
  deviations relevant to flanks of gear teeth}.
\newblock International Organization for Standardization, Geneva, 2013.
\newblock Adopted as ANSI/AGMA ISO 1328-1-B14, reaffirmed 2024.

\bibitem[{ISO}(2024)]{iso21771}
{ISO}.
\newblock {ISO 21771-1:2024, Cylindrical involute gears and gear pairs --- Part
  1: Concepts and geometry}.
\newblock International Organization for Standardization, Geneva, 2024.
\newblock First edition, 2024-10.

\bibitem[{J\"onk\"oping University, School of Engineering}(n.d.)]{ju_gears}
{J\"onk\"oping University, School of Engineering}.
\newblock {Applied Mechanics, 9.3: Modelling spur gears in {SolidWorks}}.
\newblock
  \url{https://mechanics.ju.se/MachineElements/spur-gears-solidworks.html},
  n.d.
\newblock Reports the SolidWorks Toolbox flank as a circular arc of radius
  $d/5$ centred on the base circle through the pitch point. Accessed
  2026-11-21.

\bibitem[Litvin(1998)]{litvin1998}
Faydor~L. Litvin.
\newblock Development of gear technology and theory of gearing.
\newblock NASA Reference Publication 1406 (ARL-TR-1500), NASA Lewis Research
  Center and U.S. Army Research Laboratory, 1998.
\newblock Sec.~1.1, p.~1: de La Hire, Poncelet and Camus credited with
  developing cycloidal gearing.

\bibitem[Litvin and Fuentes(2004)]{litvin2004}
Faydor~L. Litvin and Alfonso Fuentes.
\newblock \emph{Gear Geometry and Applied Theory}.
\newblock Cambridge University Press, 2nd edition, 2004.

\bibitem[Moore et~al.(2009)Moore, Kearfott, and Cloud]{moore2009}
Ramon~E. Moore, R.~Baker Kearfott, and Michael~J. Cloud.
\newblock \emph{Introduction to Interval Analysis}.
\newblock SIAM, 2009.

\bibitem[Oberg et~al.(2020)Oberg, Jones, Horton, and Ryffel]{machinery31}
Erik Oberg, Franklin~D. Jones, Holbrook~L. Horton, and Henry~H. Ryffel.
\newblock \emph{Machinery's Handbook}.
\newblock Industrial Press, South Norwalk, CT, 31st edition, 2020.
\newblock Spur Gearing, p.~2215: the $14\tfrac12^\circ$ full-depth, $20^\circ$
  stub and $14\tfrac12^\circ$ composite forms of ASA B6.1-1932, listed in ANSI
  B6.1-1968 as not recommended for new designs.

\bibitem[O'Connor and Robertson(2007)]{mactutorsavary}
John~J. O'Connor and Edmund~F. Robertson.
\newblock {F\'elix Savary}.
\newblock MacTutor History of Mathematics, University of St Andrews,
  \url{https://mathshistory.st-andrews.ac.uk/Biographies/Savary/}, 2007.
\newblock Last updated July 2007. Accessed 2026-11-21.

\bibitem[Prestridge(2023)]{prestridge2023}
Andrew Prestridge.
\newblock {Pressure Angle Standard for Involute Gearing -- 200 YEARS OLD!}
\newblock Evolvent Design, YouTube video, \url{https://youtu.be/NU-U64anuoI},
  2023.
\newblock Published 18 May 2023; history chapter at 1:12--3:50.

\bibitem[Prestridge(2026)]{involutevectors2026}
Andrew Prestridge.
\newblock Test vectors for the involute function and its inverse in {IEEE} 754
  double precision.
\newblock Dataset, Zenodo, 2026.
\newblock URL \url{https://doi.org/10.5281/zenodo.24481106}.
\newblock Licensed CC BY 4.0.

\bibitem[Radzevich(2019)]{radzevich2019}
Stephen~P. Radzevich.
\newblock Principal accomplishments in the scientific theory of gearing.
\newblock \emph{MATEC Web of Conferences}, 287:\penalty0 01001, 2019.
\newblock \doi{10.1051/matecconf/201928701001}.

\bibitem[Sibidanov et~al.(2022)Sibidanov, Zimmermann, and Glondu]{coremath2022}
Alexei Sibidanov, Paul Zimmermann, and St\'ephane Glondu.
\newblock The {CORE-MATH} project.
\newblock In \emph{Proc. 29th IEEE Symposium on Computer Arithmetic (ARITH
  2022)}, 2022.

\bibitem[{The mpmath development team}(2026)]{mpmath}
{The mpmath development team}.
\newblock mpmath: a {P}ython library for arbitrary-precision floating-point
  arithmetic (version 1.4.1).
\newblock \url{https://mpmath.org/}, 2026.

\bibitem[Volkov and Babichev(2014)]{volkov2014}
Andrey~E. Volkov and Dmitry~T. Babichev.
\newblock History of gearing theory development.
\newblock In \emph{25th Working Meeting of the IFToMM Permanent Commission for
  Standardization of Terminology on MMS}, pages 71--102, Saint Petersburg,
  2014.

\bibitem[Willis(1841)]{willis1841}
Robert Willis.
\newblock \emph{Principles of Mechanism, Designed for the Use of Students in
  the Universities, and for Engineering Students Generally}.
\newblock John W. Parker; J. \& J. J. Deighton, London; Cambridge, 1841.
\newblock Arts.~137--138, pp.~125--127: single-arc teeth with obliquity
  $75^\circ30'$, $\cos 75^\circ30'\approx1/4$; Arts.~139--142, pp.~127--132:
  two-arc teeth and the odontograph, with $75^\circ$.

\bibitem[Ziv(1991)]{ziv1991}
Abraham Ziv.
\newblock Fast evaluation of elementary mathematical functions with correctly
  rounded last bit.
\newblock \emph{ACM Transactions on Mathematical Software}, 17\penalty0
  (3):\penalty0 410--423, 1991.

\end{thebibliography}
